\documentclass[aps,prd,twocolumn,showpacs,nofootinbib]{revtex4-1}
\usepackage{graphicx}
\graphicspath{{figures/}{fig/}}
\usepackage{amsmath}
\usepackage{bm}
\usepackage{slashed}
\usepackage{epsfig}
\usepackage{amsfonts}
\usepackage{epstopdf}
\usepackage{color}
\usepackage{extarrows}
\usepackage{multirow}
\usepackage[utf8]{inputenc}
\usepackage{hyperref}

\newcommand{\ben}{\begin{eqnarray}}
\newcommand{\een}{\end{eqnarray}}

\newcommand{\bef}{\begin{figure}[!htp]}
\newcommand{\eef}{\end{figure}}

\newcommand{\bea}{\begin{eqnarray}}
\newcommand{\eea}{\end{eqnarray}}

\def\ba{\begin{linenomath*}\begin{equation}}
\def\ea{\end{equation}\end{linenomath*}}

\allowdisplaybreaks

\newcommand{\state}[4]{{^{#1}\hspace{-0.6mm}#2_{#3}^{[#4]}}}
\newcommand{\stateprime}[4]{{^{#1}\hspace{-0.6mm}#2_{#3}^{\prime[#4]}}}

\newcommand\CScSa{\state{3}{S}{1}{1}}
\newcommand\CScPz{\state{3}{P}{0}{1}}
\newcommand\CScPa{\state{3}{P}{1}{1}}
\newcommand\CScPb{\state{3}{P}{2}{1}}
\newcommand\CScPj{\state{3}{P}{J}{1}}
\newcommand\COaSz{\state{1}{S}{0}{8}}

\newcommand\COcSa{\state{3}{S}{1}{8}}
\newcommand\COcPz{\state{3}{P}{0}{8}}

\newcommand\COcPj{\state{3}{P}{J}{8}}

\newcommand{\initialstate}[2]{{#1^{[#2]}}}
\newcommand\vone{\initialstate{v}{1}}
\newcommand\veight{\initialstate{v}{8}}
\newcommand\aone{\initialstate{a}{1}}

\newcommand\lrd{\overleftrightarrow{D}}

\begin{document}

\title{   Resolving negative cross section of quarkonium hadroproduction using soft gluon factorization}

\author{An-Ping Chen$^{a}$}
\email{chenanping@jxnu.edu.cn}
\author{Yan-Qing Ma$^{b,c,d}$}
\email{yqma@pku.edu.cn}
\author{Ce Meng$^{b}$}
\email{mengce75@pku.edu.cn}

\address{$^a$ College of Physics and Communication Electronics, Jiangxi Normal University, Nanchang 330022, China\\
$^b$ School of Physics and State Key Laboratory of Nuclear Physics and Technology, Peking University, Beijing 100871, China\\
$^c$ Center for High Energy physics, Peking University,
Beijing 100871, China\\
$^d$ Collaborative Innovation Center of Quantum Matter,
Beijing 100871, China}

\date{\today}

\begin{abstract}
It was found that, using nonrelativistic QCD factorization, the predicted $\chi_{cJ}$ hadroproduction cross section at large $p_T$ can
be negative. The negative cross sections originate from terms proportional to  plus function in $\CScPj$ channels, which are remnants of the infrared subtraction in matching the $\CScPj$ short-distance coefficients. In this article, we find that the above terms can be factorized into the nonperturbative $\COcSa$ soft gluon distribution function in the soft gluon factorization (SGF) framework. Therefore, the problem can be naturally resolved in SGF. With an appropriate choice of  nonperturbative parameters, the SGF can indeed give positive predictions for $\chi_{cJ}$ production rates within the whole $p_T$ region. The production of $\psi(2S)$ is also discussed, and there is no negative cross section problem.
\end{abstract}

\maketitle
\allowdisplaybreaks

%%%%%%%%%%%%%%%%%%%%%%%%%%%%%%%%%%%%%%%%%%%%%%%%%
\section{Introduction}
Our current understanding of the inclusive heavy quarkonium production mechanism relies primarily on the non-relativistic QCD (NRQCD) factorization \cite{Bodwin:1994jh}, which
factorizes the inclusive cross section of a heavy
quarkonium into summation of perturbatively calculable short-distance coefficients
(SDCs) multiplied by nonperturbative long-distance matrix elements (LDMEs). Over the past decade, much theoretical effort has been devoted to computing the next-to-leading-order (NLO) QCD corrections to the SDCs and determining the NRQCD LDMEs. With full NLO SDCs and properly fitted LDMEs, NRQCD has been very successful in describing the yield of quarkonium states produced at the Tevatron and the LHC, including the $J/\psi$, $\psi(2S)$, $\chi_{cJ}$, and $\Upsilon(nS)$ states \cite{Ma:2010yw,Butenschoen:2010rq,Ma:2010jj,Ma:2010vd,Wang:2012is,Gong:2013qka,
Bodwin:2014gia,Shao:2014yta,Han:2014kxa,Zhang:2014coi,Bodwin:2015iua,Feng:2020cvm,
Butenschoen:2022qka}.

However, NRQCD factorization still encounters challenges in describing the inclusive quarkonium production. In addition to the well-known polarization puzzle and universality problem \cite{Campbell:2007ws, Artoisenet:2007xi,Ma:2008gq,Gong:2009kp, Zhang:2009ym,
Gong:2008sn, Gong:2008hk,Li:2011yc, Butenschoen:2012px, Chao:2012iv,
Gong:2012ug, Butenschoen:2014dra, Han:2014jya, Zhang:2014ybe,Feng:2018ukp,Chen:2022qli}, the negative cross section problem has recently emerged as a new challenge. Some studies~\cite{CHS:talk} have reported that the NRQCD predictions for charmonium yields at the LHC may turn negative when $p_T$ is very large, which is unphysical. It should be noted that this negative cross section problem is distinct from the issue of negative $p_T$-integrated quarkonium production cross sections discussed in Refs. \cite{Schuler:1994hy,Mangano:1996kg,Ozcelik:2019qze,
Lansberg:2020ejc,ColpaniSerri:2021bla,Lansberg:2021vie}. Let us consider the $p_T$ distribution of $\chi_{cJ}$ hadroproduction. It is well known that NLO contributions of $\CScPj$ for $\chi_{cJ}$ production are large but with negative signs. The negative values are originated from the remnants of the infrared subtraction in matching the $\CScPj$ SDCs. To describe experimental data, substantial cancellations between the contributions of the $\COcSa$ and $\CScPj$ channels are needed \cite{Ma:2010vd,Bodwin:2015iua}. Such cancellation will cause the $\chi_{cJ}$ production rates to become negative at exceptionally high $p_T$, as we will see in later discussions. In the region where $p_T$ is comparable to the center-of-mass energy $\sqrt{s}$, the $p_T$ behavior of $\CScPj$ contributions is sensitive to the remnant terms of the infrared subtraction. Therefore, to overcome the substantial cancellation and the negative cross section problems, one viable approach could be to modify the infrared subtraction scheme, which amounts to resum a series of relativistic correction terms comparing with the NRQCD method. This is precisely what soft gluon factorization (SGF) does. SGF is a recently proposed factorization approach \cite{Ma:2017xno}, which is equivalent to the NRQCD factorization, but with a series of important relativistic corrections
originated from kinematic effects resummed \cite{Chen:2020yeg}. In SGF the hadronization of the intermediate state to quarkonium is described by the soft gluon
distribution function (SGD), in which the momentum of soft radiation is kept. As a result, the infrared subtraction utilized in matching the $P$-wave short distance hard parts differs from that used in NRQCD, which is carried out at the point where the momentum for soft emissions is zero. Due to this, the negative cross section problem may be resolved or relieved in the SGF framework. To investigate this possibility, in this paper we apply the SGF to study the $\chi_{c}$ and $\psi(2S)$ production at LHC. For a comparison we will also present the results in NRQCD factorization.

The paper is structured as follows: In Sec. \ref{sec:LP-NLP}, we present the collinear factorization formula for calculating the hadronic production of quarkonium. In Sec. \ref{sec:FF-SGF}, we introduce the SGF of fragmentation functions and compute the related short distance hard parts. In Sec. \ref{sec:Ph-result}, we present our phenomenological results and related discussions. Our conclusions are summarized in Sec. \ref{sec:summary}. Finally, we present the calculation details of the short distance hard parts in SGF in appendix \ref{app:hard-part-results}.

\section{Quarkonium production in collinear factorization}\label{sec:LP-NLP}
As we will see, negative cross section can easily appear at high $p_T$ region, and thus we will be only interested in this region. When $p_T$ is large, the production cross section of a heavy quarkonium $H$ at hadron colliders can
be factorized as~\cite{Kang:2014tta,Kang:2014pya}
\begin{align}\label{eq:pqcdfac}
\mathrm{d}\sigma_{A+B\to H+X}(p) \approx &
\sum_{i,j}f_{i/A}(x_1,\mu_F)f_{j/B}(x_2,\mu_F)
 \nonumber\\
&\hspace{-3cm}   \times \Big \{ \sum_{f} D_{f\to H}(z,\mu_F)
    \otimes
\mathrm{d}{\hat{\sigma}}_{i+j\to f+X}({\hat P}/z,\mu_F)
 \nonumber\\
&\hspace{-3cm} +  \sum_{\kappa}  {\cal
D}_{[Q\bar{Q}(\kappa)]\to H}(z,\zeta,\zeta',\mu_F)
\\
&\hspace{-3cm} \otimes
\mathrm{d}{\hat{\sigma}}_{i+j\to [Q\bar{Q}(\kappa)]+X}({\hat P}(1\pm\zeta)/2z,{\hat
P}(1\pm\zeta')/2z,\mu_F)   \Big \}, \nonumber
\end{align}
where  $\sum_f$ runs over all parton flavors,  $\sum_\kappa$ runs over all possible spin and color states of the fragmenting $Q\bar{Q}$-pair,  $p$ is the momentum of the observed heavy quarkonium,  $\hat{P}^\mu = (p^+,0,\vec{0}_\perp)$ is a light like momentum whose plus component equals to the plus component of $p^\mu$,  $z$, $\zeta$ and $\zeta'$ are the light-cone momentum fractions, and $\mu_F$ is the collinear factorization scale.

The $\mathrm{d}{\hat{\sigma}}$'s in Eq.~\eqref{eq:pqcdfac} are perturbative calculable hard parts describing partonic interactions. $f_{i/A}$ is parton distribution function (PDF), $D_{f\to H}$ is the single parton fragmentation function (FF) which gives the leading power (LP) contribution in $1/p_T^2$ expansion, ${\cal D}_{[Q\bar{Q}(\kappa)]\to H}$ is the double parton FF~\cite{Kang:2014tta,Kang:2014pya} which gives the next-to-leading power (NLP) contribution.
The hard parts for producing a single parton have been computed up to NLO in $\alpha_s$~\cite{Aversa:1988vb}, and that for producing a heavy $Q\bar{Q}$ pair at leading-order (LO) have  been obtained in~\cite{Kang:2011mg,Kang:2014pya}. PDFs have been extracted from other experimental data and they are ready for our use. Finally for the FFs, the $\mu_F$ dependence is controlled by evolution equations. The evolution kernels are perturbatively calculable~\cite{Kang:2014tta}, but the input FFs at an given initial factorization scale $\mu_0 \gtrsim 2m_Q$ , where $m_Q$ is the heavy quark mass, are in principle nonperturbative.
Since $\mu_0\gg \Lambda_{\mathrm{QCD}}$, it is plausible to further factorize these input FFs, e.g., using the NRQCD factorization approach or the SGF approach.

The NRQCD factorization for FFs has been extensively studied. The SDCs for all double parton FFs have been calculated up to $\mathcal {O}(\alpha_s)$ in Refs.~\cite{Ma:2013yla,Ma:2014eja,Ma:2015yka}.
The SDCs for all single parton FFs are available up to $\mathcal {O}(\alpha_s^2)$~\cite{Beneke:1995yb,Braaten:1993mp,Braaten:1993rw,Cho:1994gb,Braaten:1994kd,Ma:1995vi,Braaten:1996rp,Braaten:2000pc,Hao:2009fa,Jia:2012qx,Bodwin:2014bia}
(see~\cite{Ma:2013yla,Ma:2014eja,Ma:2015yka} for a summary and comparison). And part of them are calculated to $\mathcal {O}(\alpha_s^3)$~\cite{Zhang:2017xoj,Braaten:1993rw,Braaten:1995cj,Bodwin:2003wh,Bodwin:2012xc, Zhang:2018mlo,Artoisenet:2018dbs,Feng:2018ulg, Zhang:2020atv,Zheng:2021ylc,Feng:2021uct}.
It is well known that in matching the LO SDCs of $\state{3}{P}{J}{1,8}$ channels in NRQCD factorization, the soft divergences that appear in the full-QCD expression for the $P$-wave fragmentation process are subtracted by the transition rate of $Q\bar Q$ state $\COcSa$ into $\state{3}{P}{J}{1,8}$. The subtraction is performed at the point where the momentum of soft radiation in the transition is zero, i.e., $z=1$. After the subtraction, some terms proportional to $1/(1-z)_+$ are remained. These terms cause the $P$-wave gluon fragmentation functions to strongly peak at $z\rightarrow 1$ and then become negative, leading to negative contributions to the cross sections. Especially, they drive the cross sections negative at rather large $p_T$, as we will see later.

In Ref.~\cite{Ma:2014svb}, one of the current authors and collaborators showed that, by including LP and NLP contributions of $1/p_T^2$ expansion, very simple LO calculation based on the formula Eq.~\eqref{eq:pqcdfac} can already reproduces the complicated NLO NRQCD results for both color-singlet and color-octet channels. As SGF is equivalent to NRQCD, the same conclusion should hold for SGF.
Following this, in present paper we consider only the LO contribution
to hadronic $\chi_{c}$ and $\psi(2S)$ production, i.e., we take LO PDFs, LO FFs (evaluated with LO kernels) and LO hard parts in Eq.~\eqref{eq:pqcdfac}. For simplify, we do not consider the contribution from the
$\CScSa$ channel in $\psi(2S)$ production as its contribution is much smaller than the theoretical uncertainties~\cite{Bodwin:2015iua}.

\section{Fragmentation functions in SGF } \label{sec:FF-SGF}
In SGF approach, according to Refs.~\cite{Ma:2017xno,Chen:2021hzo}, the FFs at scale $\mu_0$ can be factorized as
\begin{subequations}\label{eq:SGF-form}
\begin{align}
&D_{f \rightarrow H}(z,\mu_0) \nonumber\\
=&  \sum_{n,n^\prime} \int \frac{\mathrm{d}x}{x}   \hat{D}_{ f \to Q\bar{Q}[nn^\prime] }(\hat{z}; M_H/x, m_Q,\mu_0, \mu_\Lambda)
\nonumber\\
&\times
F_{[nn^\prime] \to H}(x,M_H,m_Q,\mu_\Lambda), \\
&{\cal D}_{[Q\bar{Q}(\kappa)] \to H}(z,\zeta,\zeta^\prime,\mu_0)
\nonumber\\ =& \sum_{n,n^\prime} \int \frac{\mathrm{d}x}{x}  \hat{\cal{D}}_{ [Q\bar{Q}(\kappa)] \to Q\bar{Q}[nn^\prime] }(\hat{z}, \zeta,\zeta^\prime; M_H/x, m_Q,\mu_0,\mu_\Lambda)
\nonumber\\
&\times F_{[nn^\prime] \to H}(x,M_H,m_Q, \mu_\Lambda),
\end{align}
\end{subequations}
where $\hat z=z/x$, $\hat{D}_{ f \to Q\bar{Q}[nn^\prime] }$ and $\hat{\cal{D}}_{ [Q\bar{Q}(\kappa)] \to Q\bar{Q}[nn^\prime] }$ are the perturbatively calculable short distance hard parts that produce a $Q \bar Q$ pair with quantum numbers $n=\state{{2S+1}}{L}{J,J_z}{c}$ and $n^\prime =\stateprime{{2S^\prime+1}}{L}{J^\prime,J_z^\prime}{c^\prime}$ in the amplitude and the complex-conjugate of the amplitude, respectively. $M_H$ is the mass of heavy quarkonium $H$ which satisfies $p^2=M_H^2$.
$F_{[nn^\prime] \to H}$ is the SGD, which describes the hadronization of an intermediate $Q\bar Q$ pair into heavy quarkonium by radiate soft gluons. The SGDs are defined as
\begin{align}\label{eq:SGD-1d}
F_{[nn^\prime] \to H}(x,M_H,m_Q,\mu_\Lambda)
&= p^+\int \frac{\mathrm{d}b^-}{2\pi} e^{-ip^+  b^-/x} \nonumber\\
&\hspace{-3cm} \times \langle 0| [\bar\Psi \mathcal {K}_{n} \Psi]^\dag(0) [a_H^\dag a_H] [\bar\Psi \mathcal {K}_{n^\prime}\Psi](b^-) |0\rangle_{\textrm{S}},
\end{align}
where $x$ is the light-cone momentum fraction which defined as $x = p^+/P_c^+$, and $P_c$ is the total momentum of the intermediate $Q\bar Q$ pair. $\Psi$ stands for Dirac field of heavy quark and the subscript ``S'' means the field operators in the above definition are the operators obtained in small momentum region. In additional, we define ``S'' to select only leading power terms in $(P_c-p)^+=(1-x)P_c^+$ expansion~\cite{Chen:2021hzo}. $\mathcal {K}_{n}$ are projection
operators corresponding to the intermediate state $n$, whose explicit definition are given in Ref.~\cite{Ma:2017xno}.

In Eq.~\eqref{eq:SGF-form}, it was suggested to expanding $m_Q^2$ around $M_H^2/(4x^2)$ in the short distance hard parts~\cite{Ma:2017xno,Chen:2021hzo},
\begin{align}\label{eq:velocity-expansion}
&\hat{D}_{ f \to Q\bar{Q}[nn^\prime] }(\hat{z}; M_H/x, m_Q,\mu_0, \mu_\Lambda)
\nonumber\\ =& \sum_{i=0} \hat{D}_{ f \to Q\bar{Q}[nn^\prime] }^{(i)}(\hat{z}; M_H/x, \mu_0, \mu_\Lambda)
\biggr(m_Q^2-\frac{M_H^2}{4x^2}\biggr)^i, \nonumber\\
& \hat{\cal{D}}_{ [Q\bar{Q}(\kappa)] \to Q\bar{Q}[nn^\prime] }(\hat{z}, \zeta,\zeta^\prime; M_H/x, m_Q,\mu_0,\mu_\Lambda) \nonumber\\
=& \sum_{i=0} \hat{\cal{D}}^{(i)}_{ [Q\bar{Q}(\kappa)] \to Q\bar{Q}[nn^\prime] }(\hat{z}, \zeta,\zeta^\prime; M_H/x, \mu_0,\mu_\Lambda) \nonumber\\&\times
\biggr(m_Q^2-\frac{M_H^2}{4x^2}\biggr)^i ,
\end{align}
which defines a velocity expansion in SGF. Here we only consider the hard parts at leading order in the velocity expansion, and then we have $n=n^\prime$. For convenience, we denote
\begin{align}
[\state{{2S+1}}{L}{J,\lambda}{c}] \equiv [\state{{2S+1}}{L}{J,\lambda}{c}\state{{2S+1}}{L}{J,\lambda}{c}].
\end{align}
Similar to the definition of polarized NRQCD LDMEs presented in~\cite{Ma:2015yka}, it is convenient to define polarized SGDs as follows:
\begin{align}
& F_{[n_\lambda] \to H}(x,M_H,m_Q, \mu_\Lambda) \nonumber\\ =& \frac{1}{N_{n_\lambda}}\sum_{\vert J_z \vert = \lambda } F_{[\state{{2S+1}}{L}{J,J_z}{c}] \to H}(x,M_H,m_Q, \mu_\Lambda),
\end{align}
where $n_\lambda$ denotes $\state{{2S+1}}{L}{J,\lambda}{c}$, $\lambda=L, T, TT, \cdots$ correspond to $\vert J_z \vert=0, 1, 2, \cdots$, respectively. $N_{n_\lambda}$ is the number of polarization states for $n_\lambda$. We have~\cite{Ma:2015yka}
\begin{align}\label{eq:N-pol}
N_{\state{{3}}{S}{1,L}{8}} &=N_{\COaSz}=N_{\state{{3}}{P}{0,0}{1,8}}= N_{\state{{3}}{P}{1,L}{1,8}}= N_{\state{{3}}{P}{2,L}{1,8}}=1,\nonumber\\
N_{\state{{3}}{S}{1,T}{8}} &= N_{\state{{3}}{P}{1,T}{1,8}}= N_{\state{{3}}{P}{2,T}{1,8}}=d-2,\nonumber\\
N_{\state{{3}}{P}{2,TT}{1,8}}&=\frac{1}{2}(d-1)(d-2)-1,
\end{align}
where $d$ is the space-time dimension.
One can also define the following unpolarized SGD
\begin{align}
& F_{[\state{{3}}{S}{1}{8}] \to H}(x,M_H,m_Q, \mu_\Lambda) \nonumber\\=& \sum_{ S_z  } F_{[\state{{3}}{S}{1,S_z}{8}] \to H}(x,M_H,m_Q, \mu_\Lambda).
\end{align}

For $\chi_{c}$ and $\psi(2S)$ production, we need to calculate the short distance hard parts for $g \to Q\bar Q[\state{{3}}{S}{1,\lambda}{8}]$, $g \to Q\bar Q[\COaSz]$, $g \to Q\bar Q[\state{{3}}{P}{J,\lambda}{1,8}]$, $[Q\bar{Q}(\kappa)] \to Q\bar Q[\state{{3}}{S}{1,\lambda}{8}] $, $[Q\bar{Q}(\kappa)] \to Q\bar Q[\COaSz] $ and $[Q\bar{Q}(\kappa)] \to Q\bar Q[\state{{3}}{P}{J,\lambda}{1,8}]$ at LO, where $\kappa=v^{[1,8]}, a^{[1,8]}$.
Following the strategy for the calculation presented in Ref.~\cite{Chen:2021hzo}, we computed these short distance hard parts. The calculation details are given in appendix~\ref{app:hard-part-results}, and the obtained results are summarized as follows
\begin{subequations}\label{eq:FFHP}
\begin{align}
& \hat{D}_{g \to Q\bar Q[\state{{3}}{S}{1,T}{8}]}^{LO,(0)}( z,M_H, \mu_0, \mu_\Lambda)
\nonumber\\ =&  \frac{\pi \alpha_s}{(N_c^2-1)}\frac{8}{M_H^3}  \delta(1-z),
\\
& \hat{D}_{g \to Q\bar Q[\state{{1}}{S}{0}{8}]}^{LO,(0)}( z,M_H, \mu_0, \mu_\Lambda) \nonumber\\ =& \frac{8\alpha_s^2}{M_H^3} \frac{N_c^2-4}{2N_c(N_c^2-1)}
 \Big[(1-z)\ln[1-z] -z^2 + \frac{3}{2}z \Big],\\
& \hat D^{LO,(0)}_{g \to Q\bar{Q}[\CScPz]}(z;M_H,\mu_0,\mu_\Lambda)
\nonumber\\ =&
        \frac{32\alpha_s^2}{M_H^5 N_c} \frac{2}{9} \Big[  \frac{1}{36} z (837 - 162 z + 72 z^2 + 40 z^3 + 8 z^4)
        \nonumber\\
        & + \frac{9  }{2 } (5 - 3 z ) \ln (1-z) \Big] , \\
& \hat D^{LO,(0)}_{g \to Q\bar{Q}[\state{{3}}{P}{1,T}{1}]}(z;M_H,\mu_0,\mu_\Lambda) \nonumber\\ =&
    \frac{32 \alpha_s^2}{M_H^5 N_c} \frac{2}{27}   z (9 + 9 z^2 + 5 z^3 + z^4) , \\
& \hat D^{LO,(0)}_{g \to Q\bar{Q}[\state{{3}}{P}{1,L}{1}]}(z;M_H,\mu_0,\mu_\Lambda) \nonumber\\ =&
\frac{32\alpha_s^2}{M_H^5 N_c} \frac{1}{27} z (9 + 18 z^2 + 10 z^3 + 2 z^4) , \\
& \hat D^{LO,(0)}_{g \to Q\bar{Q}[\state{{3}}{P}{2,TT}{1}]}(z;M_H,\mu_0,\mu_\Lambda) \nonumber\\ =&
        \frac{32\alpha_s^2}{M_H^5 N_c} \frac{2}{3z^4} \Big[ \frac{2}{9}z (108 - 216 z + 333 z^2 - 225 z^3 + 72 z^4
        \nonumber\\
        &  + 9 z^6 + 5 z^7
         + z^8)- 6(z^5-6 z^4+14 z^3-16 z^2
         \nonumber\\
        & +10 z-4 ) \ln(1-z)   \Big] , \\
& \hat D^{LO,(0)}_{g \to Q\bar{Q}[\state{{3}}{P}{2,T}{1}]}(z;M_H,\mu_0,\mu_\Lambda)
\nonumber\\ =&
        \frac{32\alpha_s^2}{M_H^5 N_c} \frac{1}{3z^4} \Big[  \frac{2}{9}z (-864 + 1728 z - 1368 z^2 + 504 z^3
         \nonumber\\
        & - 27 z^4  + 9 z^6 + 5 z^7 + z^8)  - 48 (z^4-5 z^3+10 z^2
        \nonumber\\
        & -10z+4) \ln(1-z)  \Big] , \\
&\hat D^{LO,(0)}_{g \to Q\bar{Q}[\state{{3}}{P}{2,L}{1}]}(z;M_H,\mu_0,\mu_\Lambda)
\nonumber\\ =&
        \frac{32\alpha_s^2}{M_H^5 N_c} \frac{1}{9z^4} \Big[  \frac{1}{9}z (3888 - 7776 z + 4212 z^2 - 324 z^3
        \nonumber\\
        &  - 27 z^4  + 18 z^6 + 10 z^7 +
        2 z^8)  - 216(z - 2)(z - 1)^2
        \nonumber\\
        & \times  \ln(1-z)  \Big] ,\\
&\hat D^{LO,(0)}_{g \to Q\bar{Q}[\state{{3}}{P}{J,\lambda}{8}]}(z;M_H,\mu_0,\mu_\Lambda) \nonumber\\ =& \frac{N_c^2-4}{2(N_c^2-1)} \hat D^{LO,(0)}_{g \to Q\bar{Q}[\state{{3}}{P}{J,\lambda}{1}]}(z;M_H,\mu_0,\mu_\Lambda),\\
& \hat{D}^{LO,\text{(0)}}_{[Q\bar{Q}(\veight)] \to Q\bar{Q}[\state{{3}}{S}{1,L}{8}]}(z, \zeta,\zeta^\prime;
        M_H,\mu_0,\mu_\Lambda)
        \nonumber\\ =&
        \frac{1}{N_c^2-1}\frac{2}{M_H}\frac{1}{2}
        \delta(\zeta)\delta(\zeta^\prime)\delta(1-z), \\
&\hat{D}^{LO,\text{(0)}}_{[Q\bar{Q}(a^{[8]})] \to Q\bar{Q}[\COaSz]}(z, \zeta,\zeta^\prime;
        M_H,\mu_0,\mu_\Lambda)
        \nonumber\\ =& \frac{1}{(N_c^2-1)}\frac{2}{M_H}\frac{1}{2}\delta(\zeta)\delta(\zeta^\prime)
\delta(1-z),\\
& \hat{D}^{LO,\text{(0)}}_{[Q\bar{Q}(\vone)] \to Q\bar{Q}[\CScPz]}(z, \zeta,\zeta^\prime;
        M_H,\mu_0,\mu_\Lambda)
        \nonumber\\ =&
         \frac{8}{M_H^3}\frac{1}{6}\delta'(\zeta)\delta'(\zeta^\prime)\delta(1-z),
        \\
& \hat{D}^{LO,\text{(0)}}_{[Q\bar{Q}(\aone)] \to Q\bar{Q}[\state{{3}}{P}{1,L}{1}]}(z, \zeta,\zeta^\prime;
        M_H,\mu_0,\mu_\Lambda)
        \nonumber\\ =&
         \frac{8}{M_H^3} \delta(\zeta)\delta(\zeta^\prime)\delta(1-z),\\
& \hat{D}^{LO,\text{(0)}}_{[Q\bar{Q}(\vone)] \to Q\bar{Q}[\state{{3}}{P}{2,L}{1}]}(z, \zeta,\zeta^\prime;
        M_H,\mu_0,\mu_\Lambda)
        \nonumber\\ =& \frac{8}{M_H^3}\frac{1}{3}\delta'(\zeta)\delta'(\zeta^\prime)\delta(1-z),\\
& \hat{D}^{LO,\text{(0)}}_{[Q\bar{Q}(s^{[8]})] \to Q\bar{Q}[\state{{3}}{P}{J,\lambda}{8}]} (z, \zeta,\zeta^\prime;
        M_H,\mu_0,\mu_\Lambda)\nonumber\\ =&
         \frac{1}{N_c^2-1}  \hat{D}^{LO,\text{(0)}}_{[Q\bar{Q}(s^{[1]})] \to Q\bar{Q}[\state{{3}}{P}{J,\lambda}{1}]} (z, \zeta,\zeta^\prime;
        M_H,\mu_0,\mu_\Lambda),
\end{align}
\end{subequations}
where $s$ could be $v$ or $a$.
Here we only listed the nonvanishing LO hard parts. The contributions of $t^{[1,8]}$ channels for double parton FFs are neglected, due to the partonic SDCs $d{\hat{\sigma}}_{i+j\to [Q\bar{Q}(t^{[1,8]})]+X}$ vanishes at LO [$\mathcal {O}(\alpha_s^3)$]~\cite{Kang:2014pya}. In contrast to the NRQCD SDCs, the $P$-wave short distance hard parts presented above do not include terms proportional to $1/(1-z)_+$. These plus distributions correspond to the leading-power order in the momentum of the emitted soft gluons, which are factorized into the $\COcSa$ SGD in SGF.

In the SGF formula Eq.~\eqref{eq:pqcdfac}, a factorization scale $\mu_\Lambda$ is introduced. The freedom to choose $\mu_\Lambda$ results from the RG equations obeyed by the SGDs~\cite{Chen:2021hzo}. For simplify, here we do not consider the evolution of SGDs and directly provide a model for them at the scale $\mu_\Lambda=M_H$. We adopt following model~\cite{Lee:2021oqr,Cacciari:2005uk} for the SGDs
\begin{align}\label{eq:model-SGD-SW}
F^{\textrm{mod}}(x)=&  \frac{N^H \Gamma(M_Hb/\bar{\Lambda})(1-x)^{b-1} x^{M_Hb/\bar{\Lambda}-b-1}}{\Gamma(M_Hb/\bar{\Lambda}-b)
 \Gamma(b)}.
\end{align}
Here $N^H$ determines the normalization, $\bar{\Lambda}$ characterizes the average radiated momentum in the hadronization process, and $b$ is related to the second moment of model function, i.e.,
\begin{subequations}\label{eq:moments}
\begin{align}
& \int_0^1\mathrm{d}x F^{\textrm{mod}}(x)=N^H,\\
& \int_0^1\mathrm{d}x M_H(1-x) F^{\textrm{mod}}(x)=N^H \bar{\Lambda},\\
& \int_0^1\mathrm{d}x \Big(M_H(1-x)\Big)^2 F^{\textrm{mod}}(x)=\frac{N^H \bar{\Lambda}^2(b+1)}{b+\bar{\Lambda}/M_H},
\end{align}
\end{subequations}
where $M_H(1-x)$ denotes the radiated momentum in the hadronization, and $\bar{\Lambda}$ should be $\mathcal{O}(\Lambda_{\textrm{QCD}})$.
 These parameters are depend on the $Q\bar Q$ state $n$. The normalization factor $N^H[n]$ is related to the NRQCD LDME $\langle\mathcal {O} ^H(n)\rangle$ defined in~\cite{Ma:2015yka}. Up to the lowest order in velocity expansion, we have~\cite{Ma:2017xno}
\begin{align}
N^H[n]\approx \langle\mathcal {O} ^H(n)\rangle.
\end{align}
 Similar to the LDMEs in NRQCD, we assume that these normalization factors satisfy the spin symmetry relations
\begin{align}
N^{\chi_{c0}}[\COcSa] &=3 N^{\chi_{c0}}[\state{{3}}{S}{1,T}{8}]=3N^{\chi_{c0}}[\state{{3}}{S}{1,L}{8}]\nonumber\\
  &=  N^{\chi_{c1}}[\state{{3}}{S}{1,T}{8}]=N^{\chi_{c1}}[\state{{3}}{S}{1,L}{8}] \nonumber\\
  &=  \frac{3}{5} N^{\chi_{c2}}[\state{{3}}{S}{1,T}{8}] =  \frac{3}{5} N^{\chi_{c2}}[\state{{3}}{S}{1,L}{8}], \nonumber\\
N^{\chi_{cJ}}[\state{{3}}{P}{J,\lambda}{1}] &= N^{\chi_{c0}}[\CScPz] , \nonumber\\
N^{\psi(2S)}[\COcSa] &=3 N^{\psi(2S)}[\state{{3}}{S}{1,T}{8}]=3N^{\psi(2S)}[\state{{3}}{S}{1,L}{8}] , \nonumber\\
N^{\psi(2S)}[\state{{3}}{P}{J,\lambda}{8}] &= N^{\psi(2S)}[\COcPz] .
\end{align}
These relations reduce the number of normalization factors required, and we select $N^{\chi_{c0}}[\COcSa]$, $N^{\chi_{c0}}[\CScPz]$, $N^{\psi(2S)}[\COcSa]$, $N^{\psi(2S)}[\COaSz]$, and $N^{\psi(2S)}[\COcPz]$ as the free parameters, which can be determined through fitting to experimental data.

\section{Phenomenological results} \label{sec:Ph-result}

\subsection{General setup }\label{sec:setup}

Before going ahead, we provide some details regarding our fitting procedure. We use LO PDFs, LO SDCs and LO FFs in Eq.~\eqref{eq:pqcdfac}. We use CTEQ6L1 \cite{Pumplin:2002vw} as the input LO PDF, and use LO $\alpha_s$ with $n_f = 5$ and $\Lambda^{(5)}_{\textrm{QCD}}=165~\rm{MeV}$. The LO [$\mathcal{O}(\alpha_s^2)$]
SDCs for LP are known in \cite{Aversa:1988vb}, and the complete LO [$\mathcal{O}(\alpha_s^3)$] SDCs for NLP are taken from Ref.~\cite{Kang:2011mg,Kang:2014pya}. To facilitate comparison, we evaluate the LO FFs in both the NRQCD factorization and the SGF approaches. We fix the charm quark mass to $m_Q=m_c=1.5~\rm{GeV}$ and choose the factorization scale in Eq.~\eqref{eq:pqcdfac} to be $\mu_F=p_T$. To resum the leading logarithms of $p_T^2/m_c^2$,
we evolve the single parton FFs from the initial scale $\mu_0$ to the scale $\mu_F=p_T$ \footnote{Here we do not include the evolution of double parton FFs, because the solution of the corresponding RGEs is still absent. Nevertheless, this should be tolerable because logarithms are important only in the region where $p_T$ is very large, but then the double parton contribution is suppressed comparing with the single parton contribution.}. We compute the evolved FFs by using the method described in Ref.~\cite{Bodwin:2015iua}. In the calculation, we use the LO evolution kernel with $n_f=3$. In NRQCD factorization, we choose initial scale $\mu_0=2m_c$ and set NRQCD factorization scale $\mu_\Lambda=m_c$ as usual. In SGF, we take $\mu_0=M_H$ and set SGF factorization scale $\mu_\Lambda=M_H$. For $H=\chi_{cJ}$, we choose $M_{H} = 3.5~\rm{GeV}$, and for $H=\psi(2S)$ we choose $M_{H} = m_{\psi(2S)} =3.686~\rm{GeV}$~\cite{ParticleDataGroup:2010dbb}.

We use the data on the $\chi_{c1}$ and $\chi_{c2}$ transverse
momentum distributions provided by ATLAS Collaboration~\cite{ATLAS:2014ala} at $\sqrt{s}=7~\rm{TeV}$ to perform the fits of the parameters $N^{\chi_{c0}}[\COcSa]$, $N^{\chi_{c0}}[\CScPz]$ in SGF, as well as the NRQCD LDMEs
$\langle \mathcal{O}^{\chi_{c0}}(\COcSa)\rangle$ and $\langle \mathcal{O}^{\chi_{c0}}(\CScPz)\rangle$. We determine the parameters and LDMEs for $\psi(2S)$ by fitting to the cross section data provided by ATLAS and CMS Collaborations~\cite{ATLAS:2014zpz,CMS:2015lbl}. Due to the factorization formula Eq.~\eqref{eq:pqcdfac} is applied in the large $p_T$ region, only the data with $p_T\geq 12\rm{GeV}$ are considered in the fitting. Following Ref.~\cite{Bodwin:2015iua}, in the fitting we take the uncertainties in the theoretical expressions
for the charmonium cross sections to
be $30\%$ of the central values in magnitude, which
account for uncalculated corrections of higher orders in
velocity expansion. Some branching ratios are used in the fits, they are $B(\chi_{c0} \to J/\psi \gamma)=0.0128$, $B(\chi_{c1} \to J/\psi \gamma)=0.36$, $B(\chi_{c2} \to J/\psi \gamma)=0.2$, $B(\psi(2S) \to \mu^+\mu^- )=0.0075$, $B(\psi(2S) \to J/\psi\pi^+\pi^- )=0.34$, $B(J/\psi \to \mu^+\mu^- )=0.0593$~\cite{ParticleDataGroup:2010dbb}.

\subsection{Production of $\chi_{cJ}$}\label{sec:result}

\begin{figure}[htb!]
 \includegraphics[width=0.8\linewidth]{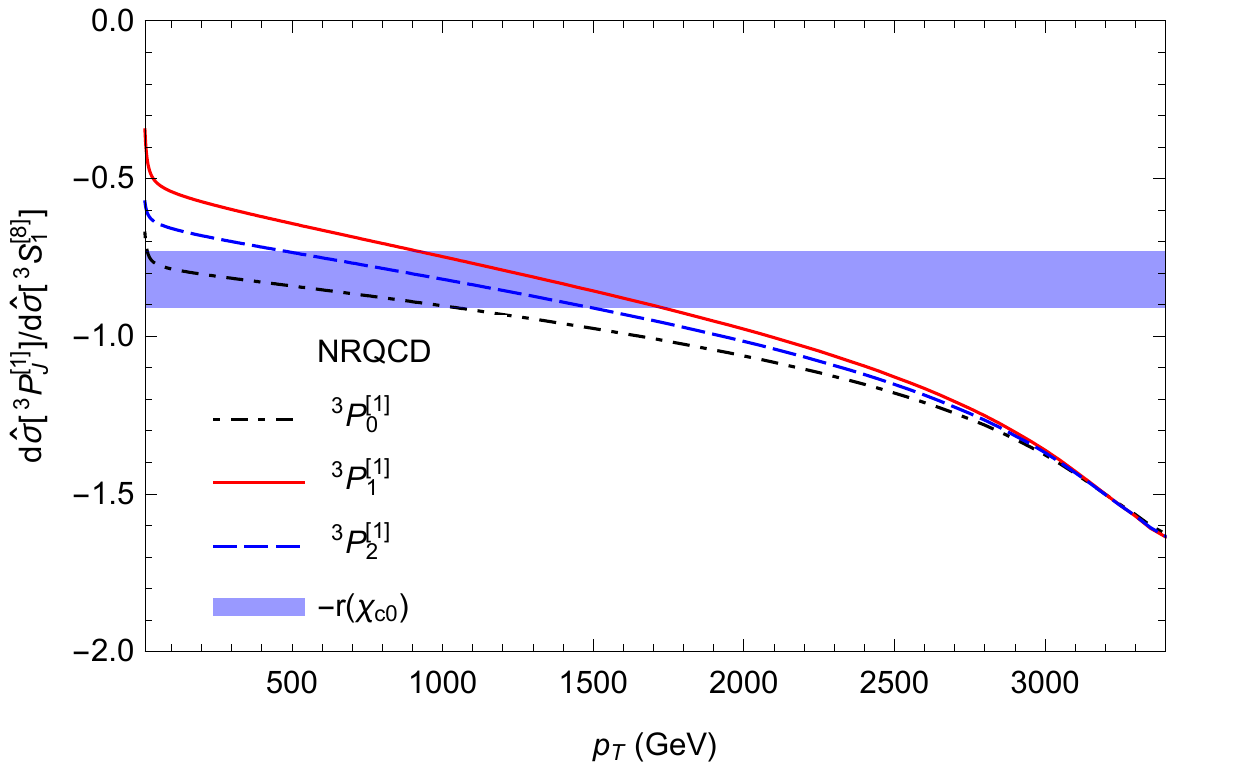}
 \includegraphics[width=0.8\linewidth]{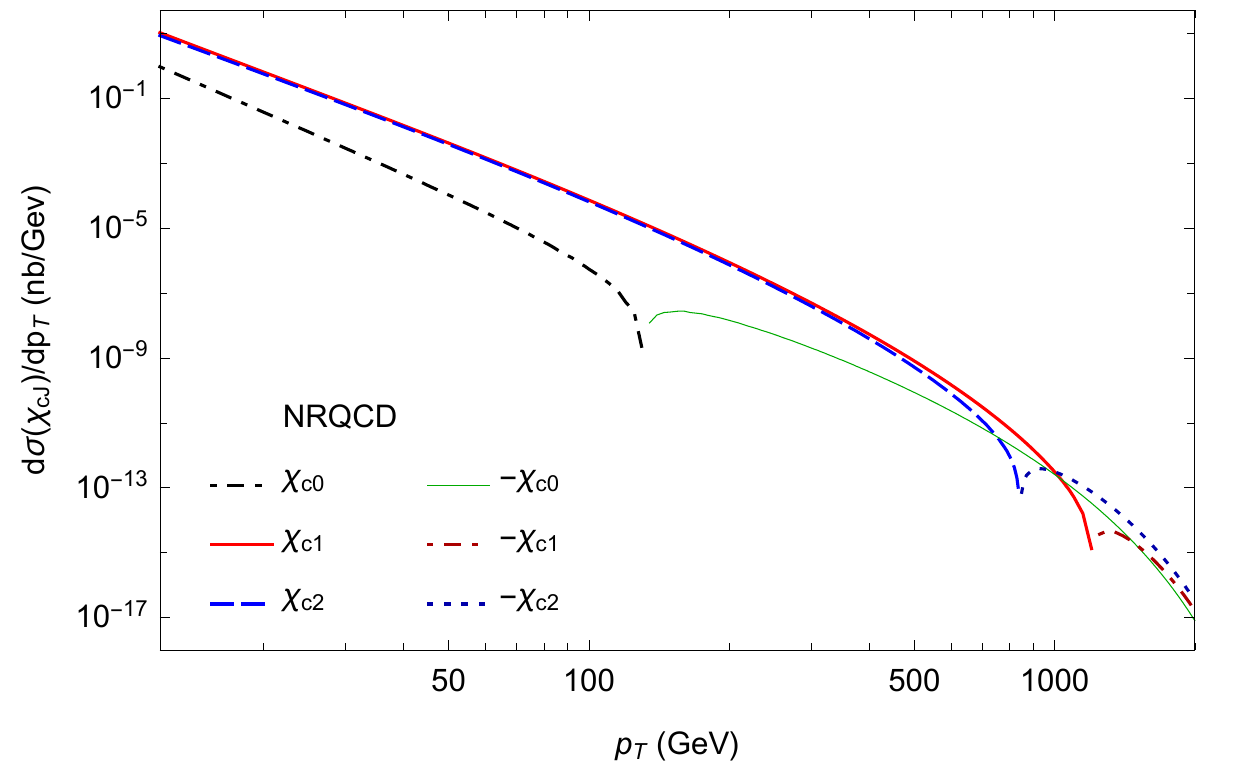}
 \caption{Upper panel: the comparison between the ratios $d\hat{\sigma}[\CScPj]/d\hat{\sigma}[\COcSa]$ and $-r(\chi_{c0})$. Lower panel: the $p_T$ distributions for $\chi_{cJ}$ production when the LDMEs take the central values in Eq.~\eqref{eq:NREM}. Here $\sqrt{s}=7\rm{TeV}$, $|y|<0.75$. }\label{fig:NRratio}
\end{figure}

In NRQCD factorization, through the least-$\chi^2$ fit to
the measured data, we obtain
\begin{subequations}\label{eq:NREM}
\begin{align}
\langle \mathcal{O}^{\chi_{c0}}(\COcSa)\rangle=(4.84\pm 1.14)\times 10^{-3} \textrm{GeV}^3,\\
\frac{\langle \mathcal{O}^{\chi_{c0}}(\CScPz)\rangle}{m_c^2}=(6.11\pm 2.01)\times 10^{-3} \textrm{GeV}^3,
\end{align}
\end{subequations}
with $\chi^2/\textrm{d.o.f}=0.53/8$. The above values of NRQCD LDMEs are consistent with the those obtained in Ref.~\cite{Bodwin:2015iua}.
%%%%%%%%%%%%%%%%%%%%%%%%%%%%%%%%%%%%%%%%%%%%%%%%%%%%%%%%%%%%%%%%%%%%%%%%%%%%%%%%%%%%%%%
\begin{figure}[htb!]
 \includegraphics[width=0.8\linewidth]{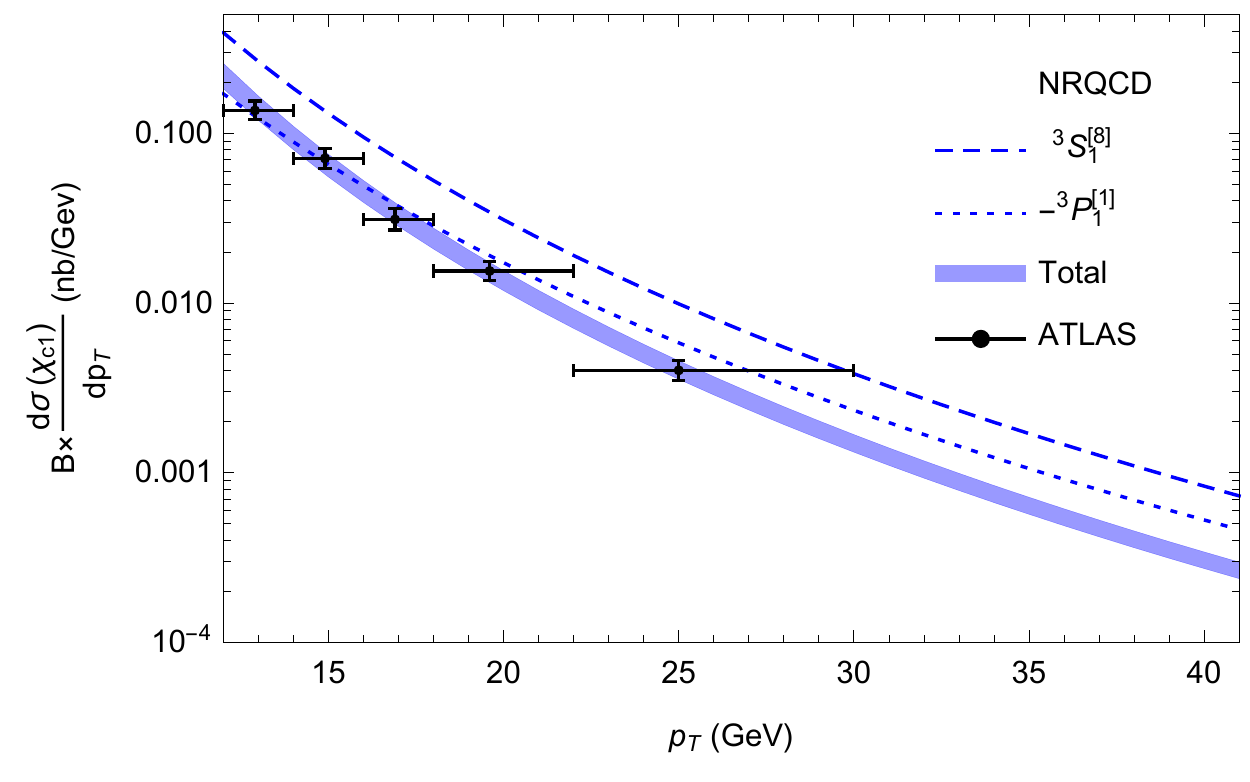}
 \includegraphics[width=0.8\linewidth]{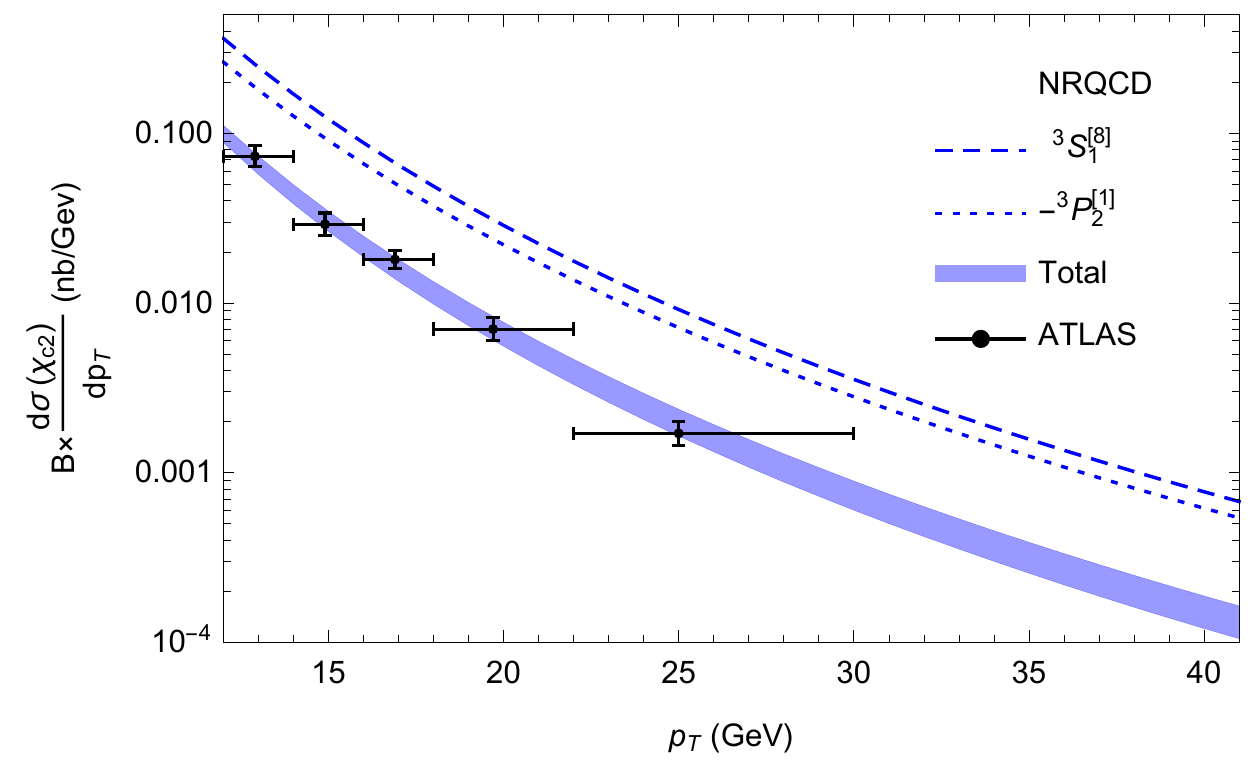}
 \caption{The differential cross sections for prompt $\chi_{c1}$ and $\chi_{c2}$ at the LHC center of mass energy $\sqrt{s}=7~\rm{TeV}$ and in the rapidity range $|y|<0.75$ compared with ATLAS measurements~\cite{ATLAS:2014ala}. $B=B(\chi_{cJ} \to J/\psi \gamma) \times B(J/\psi \to \mu^+\mu^- )$. }\label{fig:NRatlas}
\end{figure}
The $\chi_{c1}$ and $\chi_{c2}$ cross sections obtained from the fit are depicted in  Fig.~\ref{fig:NRatlas} against ATLAS data~\cite{ATLAS:2014ala}, where the contributions of the individual channels to the cross sections are also presented. It is evident that significant cancellations occur between the contributions of the $\COcSa$ and $\CScPj$ channels in NRQCD factorization. This kind of cancellation can make the perturbative expansion unstable and may even lead to negative cross sections at large $p_T$. To see this more clear, we define the ratio
\begin{align}
r(\chi_{c0})\equiv\frac{\langle \mathcal{O}^{\chi_{c0}}(\COcSa)\rangle}{\langle \mathcal{O}^{\chi_{c0}}(\CScPz)\rangle/m_c^2},
\end{align}
and rewrite the cross sections as
\begin{align}\label{eq:NRCS}
d\sigma(\chi_{cJ})=& (2J+1) d\hat{\sigma}[\COcSa] \frac{
\langle \mathcal{O}^{\chi_{c0}}(\CScPz)\rangle }{ m_c^2 } \nonumber\\
  &\times \biggr[ r(\chi_{c0}) + \frac{d\hat{\sigma}[\CScPj]}{d\hat{\sigma}[\COcSa]} \biggr].
\end{align}
According to the fit we have $r(\chi_{c0})=0.79^{+0.12}_{-0.06}$.
To achieve a positive production rate of $\chi_{cJ}$ at high $p_T$, it is necessary to have
\begin{align}
\frac{d\hat{\sigma}[\CScPj]}{d\hat{\sigma}[\COcSa]}> -r(\chi_{c0}).
\end{align}
However, as showed in the upper panel of Fig.~\ref{fig:NRratio}, we find the ratios $d\hat{\sigma}[\CScPj]/d\hat{\sigma}[\COcSa](J=0,1,2)$ decrease as $p_T$ increases. Moreover, for $p_T$ larger than about $1700 ~\textrm{GeV}$, all three ratios fall below the lower bound of $-r(\chi_{c0})$. Consequently, at large $p_T$ all cross sections become negative, as showed in the lower panel of Fig.~\ref{fig:NRratio}, which has also been pointed out in Ref.~\cite{CHS:talk}.

It is worth noting that the cross section in the very high $p_T$ region is greatly influenced by the behavior of fragmentation functions near $z=1$.
In this regime, the $p_T$ behavior of $\COcSa$ channel is mainly dominated by the delta function $\delta(1-z)$, while that of $\CScPj$ channel is primarily governed by the terms proportional to $1/(1-z)_+$. The absolute value of the latter decreases more slowly as $p_T$ increases compared to the former, leading to an increase in the ratios $-d\hat{\sigma}[\CScPj]/d\hat{\sigma}[\COcSa]$.
%%%%%%%%%%%%%%%%%%%%%%%%%%%%%%%%%%%%%%%%%%%%%%%%%%%%%%%%%%%%%%%%%%%%%%%%%%%%%%%%%%%%%%%

\begin{figure*}
 \includegraphics[width=1\linewidth]{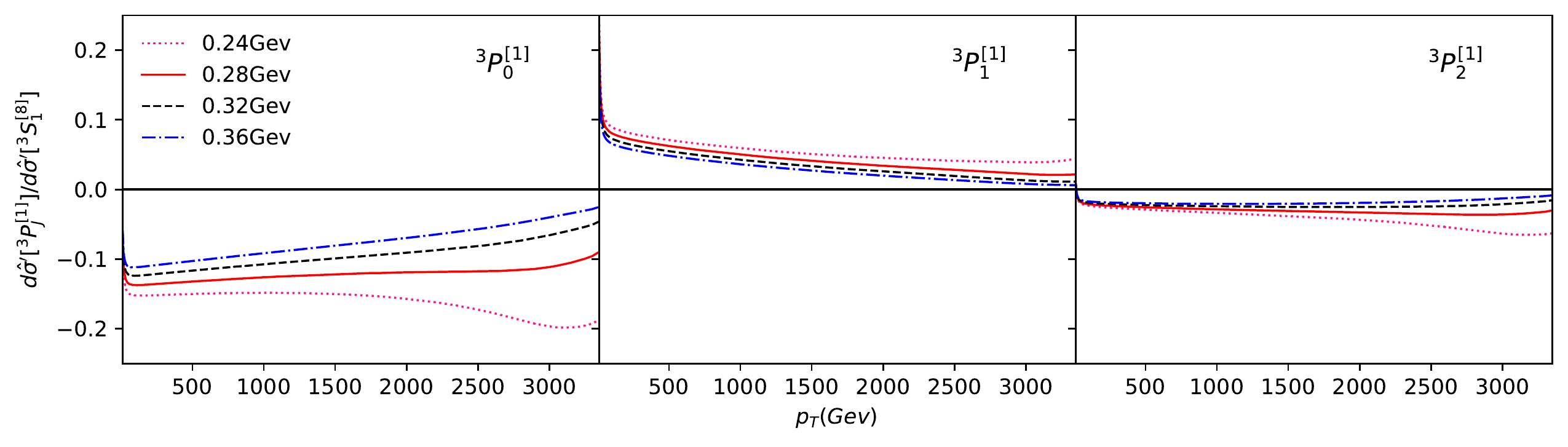}
 \caption{$\bar{\Lambda}[\CScPj]$ dependence of the $p_T$ distribution for the ratio $d\hat{\sigma}^\prime[\CScPj]/d\hat{\sigma}^\prime[\COcSa]$. Here we fix $\bar{\Lambda}[\COcSa]=0.4\textrm{GeV}$ and vary $\bar{\Lambda}[\CScPj]=0.36$, $0.32$, $0.28$, $0.24\textrm{GeV}$. Here $\sqrt{s}=7\rm{TeV}$, $|y|<0.75$.} \label{fig:SGFratio1}.
\end{figure*}

\begin{figure*}
 \includegraphics[width=1\linewidth]{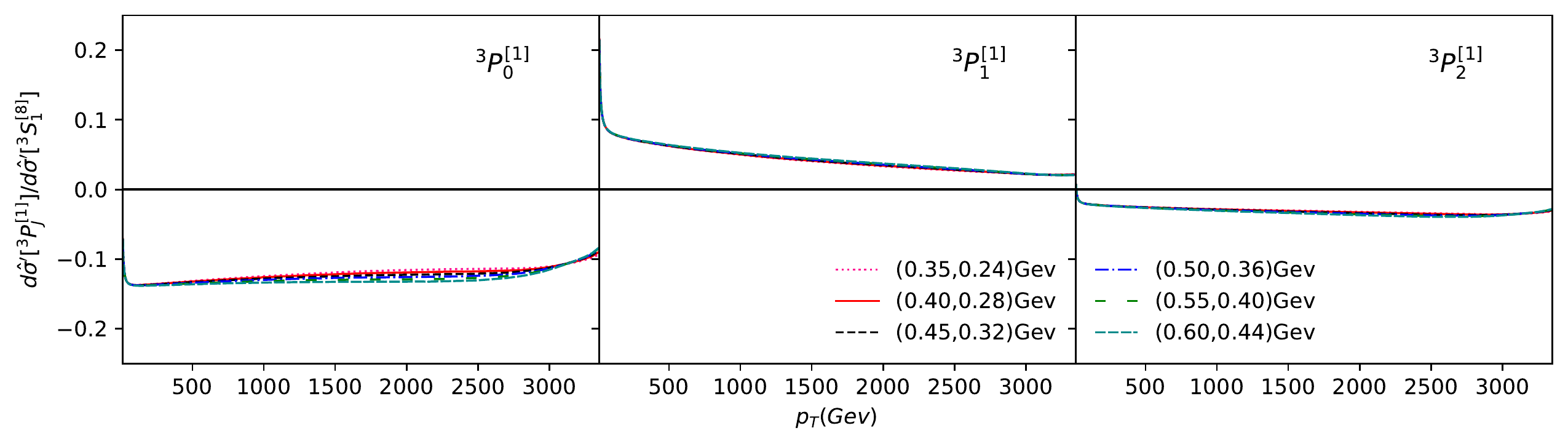}
 \caption{The $p_T$ distribution for the ratio $d\hat{\sigma}^\prime[\CScPj]/d\hat{\sigma}^\prime[\COcSa]$ with $\bar{\Lambda}[\CScPj]=\bar{\Lambda}_{\textrm{min}} \approx 0.7\bar{\Lambda}[\COcSa]$. We vary $(\bar{\Lambda}[\COcSa], \bar{\Lambda}[\CScPj])=(0.35,0.24)$, $(0.4,0.28)$, $(0.45,0.32)$, $(0.5,0.36)$, $(0.55,0.4)$, $(0.6,0.44)\textrm{GeV}$. Here $\sqrt{s}=7\rm{TeV}$, $|y|<0.75$.} \label{fig:SGFratio2}
 \vspace*{0.cm}
\end{figure*}

In SGF, after convolving SGDs with short distance hard parts in Eq.~\eqref{eq:FFHP}, we arrive at a form that similar to Eq.~\eqref{eq:NRCS},
\begin{align}\label{eq:SGFCS}
d\sigma(\chi_{cJ}) = & (2J+1) d\hat{\sigma}^\prime[\COcSa] \frac{
 N^{\chi_{c0}}[\CScPz]  }{ m_c^2 }  \nonumber\\
 & \times \biggr[ r^\prime(\chi_{c0}) + \frac{d\hat{\sigma}^\prime[\CScPj]}{d\hat{\sigma}^\prime[\COcSa]} \biggr].
\end{align}
Here
\begin{align}
r^\prime(\chi_{c0})  \equiv \frac{N^{\chi_{c0}}[\COcSa]}{N^{\chi_{c0}}[\CScPz]/m_c^2}.
\end{align}
The coefficients $d\hat{\sigma}^\prime[\COcSa]$ and $d\hat{\sigma}^\prime[\CScPj]$ depend on the nonperturbative parameters $b$ and $\bar{\Lambda}$. The behavior of FFs near $z= 1$ is dominated by the nonperturbative SGDs. Thus in the very high $p_T$ region, the $p_T$ behavior of the ratio $d\hat{\sigma}^\prime[\CScPj]/d\hat{\sigma}^\prime[\COcSa]$ is very sensitive to the parameters $b$ and $\bar{\Lambda}$. Unfortunately, there is yet no first-principle way to determine these parameters. From Eq.~\eqref{eq:moments} we find that when $b>1$, the second moment of the model is $\mathcal{O}(N^H \bar{\Lambda}^2)$ and is not sensitive to $b$.  Therefore, for simplify here we assume that
the parameter $b$ is same for all channels and we set $b=6$. To qualitatively analyze the effect of $\bar{\Lambda}$ on $d\hat{\sigma}^\prime[\CScPj]/d\hat{\sigma}^\prime[\COcSa]$, we first fix $\bar{\Lambda}[\COcSa]=0.4~\textrm{GeV}$ and vary $\bar{\Lambda}[\CScPj]$. In Fig.~\ref{fig:SGFratio1} we show the ratios for $\bar{\Lambda}[\CScPj]=0.36$, $0.32$, $0.28$, $0.24~\textrm{GeV}(J=0,1,2)$. We find that the coefficient in $\CScPa$ channel is always positive, and those in $\CScPz$ and $\CScPb$ channels are negative. Additionally, for a fixed $\bar{\Lambda}[\COcSa]$, there exist a scale $\bar{\Lambda}_{\textrm{min}}$ such that when $\bar{\Lambda}[\CScPj] \geq \bar{\Lambda}_{\textrm{min}}$, the minimum value of the ratio for $J=0$ is located at $p_T<100 \textrm{GeV}$, and when $\bar{\Lambda}[\CScPj] < \bar{\Lambda}_{\textrm{min}}$, the position of minimum value turns to be $p_T \sim 3000 \textrm{GeV}$. To obtain a positive cross section for $\chi_{cJ}$ production within the whole $p_T$ region, a suitable choice is $\bar{\Lambda}[\CScPj] \geq \bar{\Lambda}_{\textrm{min}}$. This indicates that in the hadronization of $Q\bar Q[\CScPj]$ to $\chi_{cJ}$ the average radiated momentum should not too close to zero, which is different from the picture in NRQCD factorization. On the other hand, to investigate the correlation between $\bar{\Lambda}_{\textrm{min}}$ and $\bar{\Lambda}[\COcSa]$, we vary $\bar{\Lambda}[\COcSa]=0.6$, $0.55$, $0.5$, $0.45$, $0.4$, $0.35~\textrm{GeV}$ and roughly estimate the $\bar{\Lambda}_{\textrm{min}}$ for each corresponding $\bar{\Lambda}[\COcSa]$. Our analysis suggests that $\bar{\Lambda}_{\textrm{min}} \approx 0.7\bar{\Lambda}[\COcSa]$. The ratios for each $(\bar{\Lambda}[\COcSa], \bar{\Lambda}[\CScPj])=(\bar{\Lambda}[\COcSa], \bar{\Lambda}_{\textrm{min}})$ pair are showed in Fig.~\ref{fig:SGFratio2}.
%%%%%%%%%%%%%%%%%%%%%%%%%%%%%%%%%%%%%%%%%%%%%%%%%%%%%%%%%%%%%%%%%%%%%%%%%%%%%%%%%%%%%%%
%%%%%%%%%%%%%%%%%%%%%%%%%%%%%%%%%%%%%%%%%%%%%%%%%%%%%%%%%%%%%%%%%%%%%%%%%%%%%%%%%%%%%%%
%%%%%%%%%%%%%%%%%%%%%%%%%%%%%%%%%%%%%%%%%%%%%%%%%%%%%%%%%%%%%%%%%%%%%%%%%%%%%%%%%%%%%%%
%%%%%%%%%%%%%%%%%%%%%%%%%%%%%%%%%%%%%%%%%%%%%%%%%%%%%%%%%%%%%%%%%%%%%%%%%%%%%%%%%%%%%%%

\begin{figure}[htb!]
 \begin{center}
 \includegraphics[width=0.8\linewidth]{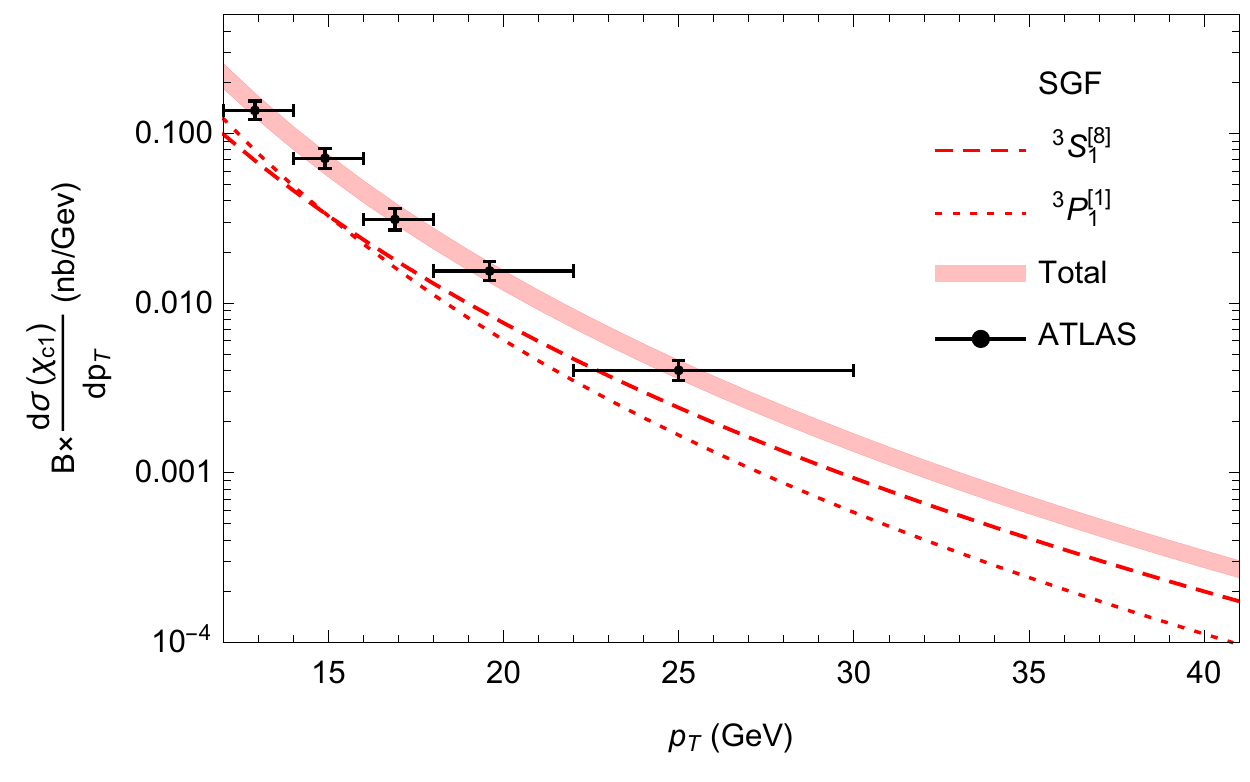}
 \includegraphics[width=0.8\linewidth]{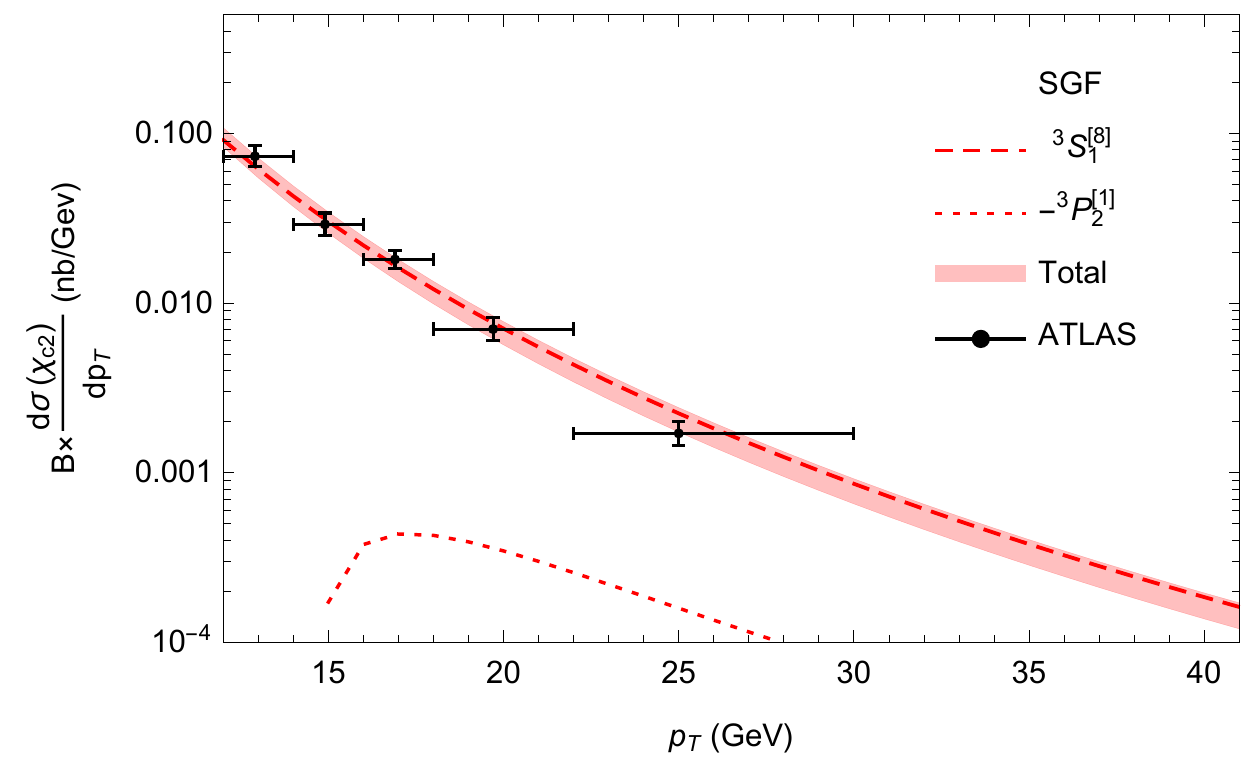}
 \end{center}
 \caption{The differential cross sections for $\chi_{c1}$ and $\chi_{c2}$ production in SGF approach compared with ATLAS measurements. Here $\sqrt{s}=7~\rm{TeV}$, $|y|<0.75$.}\label{fig:sgfatlas}
\end{figure}

\begin{figure}[htb!]
\begin{center}
 \includegraphics[width=0.8\linewidth]{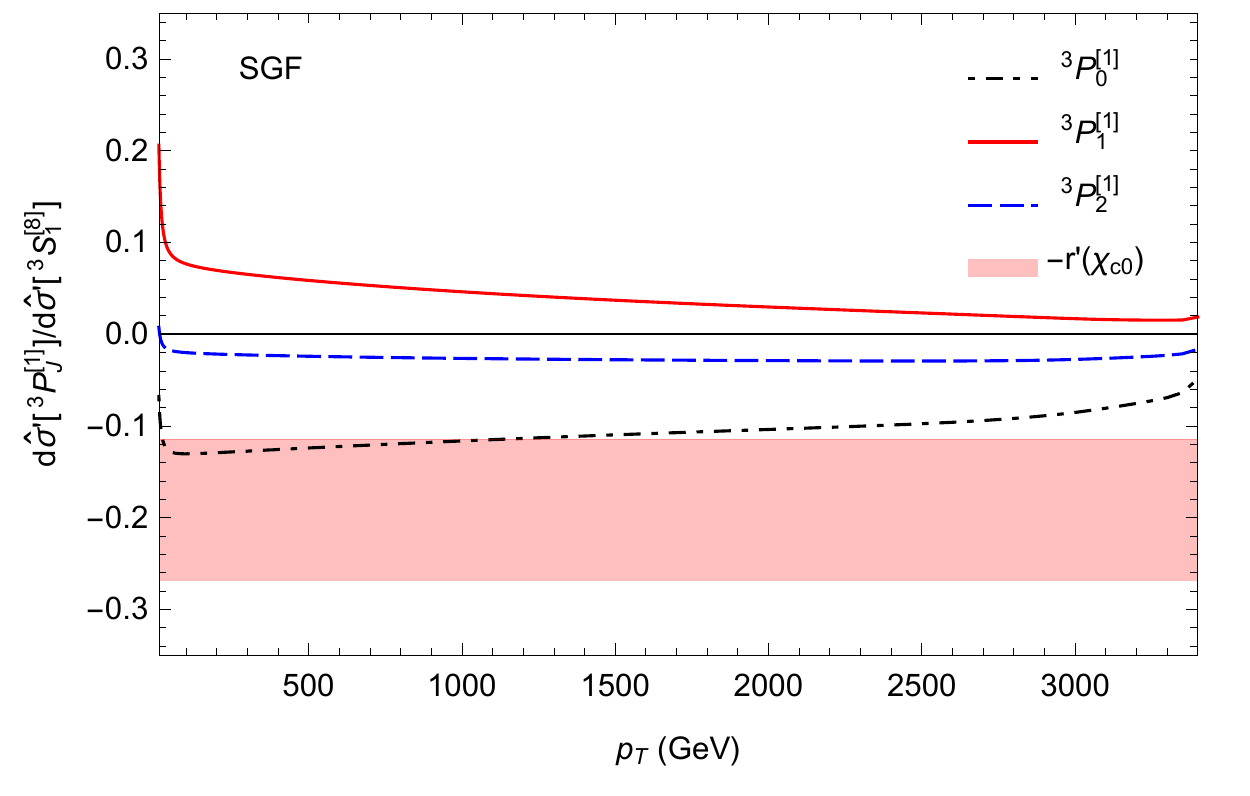}
 \includegraphics[width=0.8\linewidth]{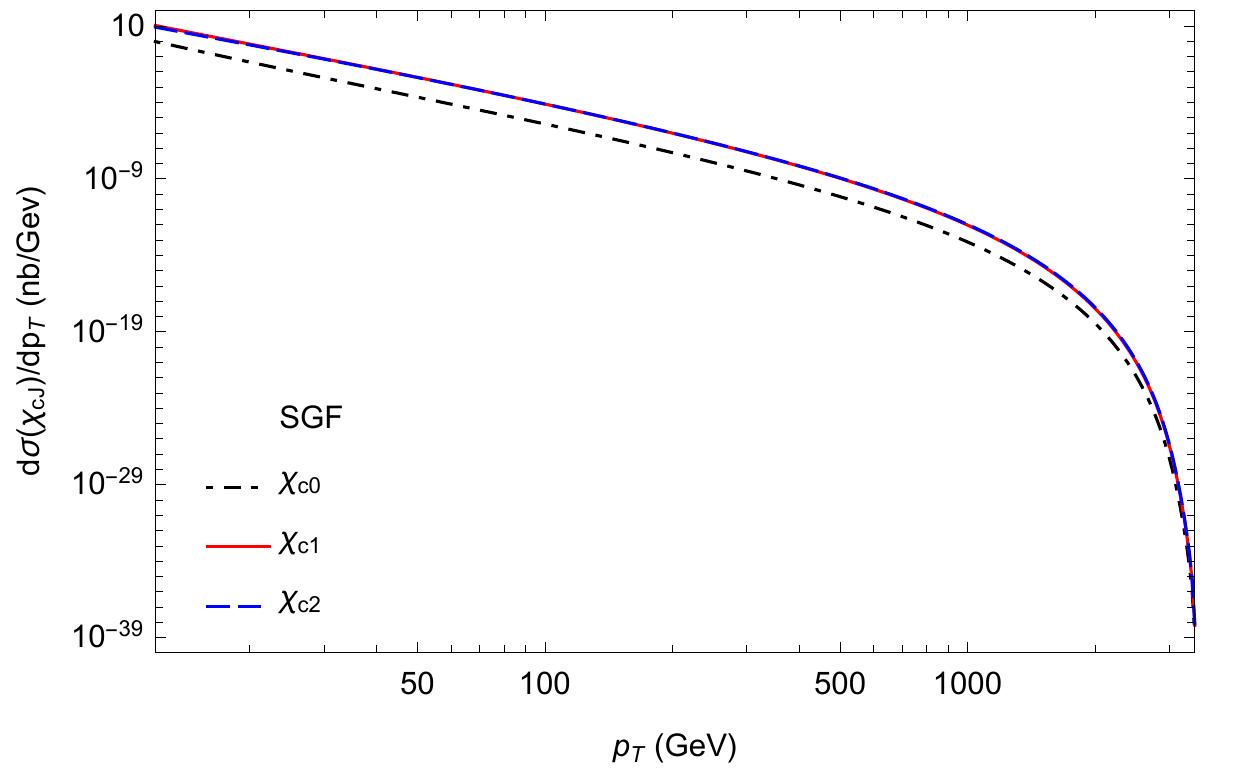}
 \end{center}
 \caption{Upper panel: the comparison between the ratios $d\hat{\sigma}^\prime[\CScPj]/d\hat{\sigma}^\prime[\COcSa]$ and $-r^\prime(\chi_{c0})$. Lower panel: the $p_T$ distributions for $\chi_{cJ}$ production with $N^{\chi_{c0}}[\COcSa]$, $N^{\chi_{c0}}[\CScPz]/m_c^2$ take the central values in Eq.~\eqref{eq:SGFLDEM}. Here $\sqrt{s}=7~\rm{TeV}$, $|y|<0.75$.}\label{fig:sgfratio}
\end{figure}

Based on the discussion above, we set $\bar{\Lambda}[\COcSa]=0.4~\textrm{GeV}$ and $\bar{\Lambda}[\CScPj]=0.3~\textrm{GeV}$ for the $\chi_{cJ}$ production. Subsequently, we determine the parameters $N^{\chi_{c0}}[\COcSa]$ and $N^{\chi_{c0}}[\CScPz]$ by fitting the experimental data, which results in
\begin{subequations}\label{eq:SGFLDEM}
\begin{align}
N^{\chi_{c0}}[\COcSa]=(3.75 \pm 0.53)\times 10^{-3} \textrm{GeV}^3,\\
\frac{N^{\chi_{c0}}[\CScPz]}{m_c^2}=(2.25\pm 0.75)\times 10^{-2} \textrm{GeV}^3,
\end{align}
\end{subequations}
with $\chi^2/\textrm{d.o.f}=0.63/8$. We also obtain $r^\prime(\chi_{c0})=0.17^{+0.10}_{-0.06}$. The fitted cross sections are shown in Fig.~\ref{fig:sgfatlas}. In SGF, there is no substantial cancellations between
the contributions of the $\COcSa$ and $\CScPj$ channels.
Moreover, the contribution of $\CScPa$ channel in the $\chi_{c1}$ cross section is always positive. We also find that in the experimental $p_T$ region, the
$\chi_{c2}$ cross section is dominated by the contribution of $\COcSa$ channel, and the color-singlet contribution can be neglected. This indicates that the parameter $N^{\chi_{c0}}[\COcSa]$ is mainly determined by the differential cross section for prompt $\chi_{c2}$ production. In the upper panel of Fig.~\ref{fig:sgfratio}, the $p_T$ dependence of ratios $d\hat{\sigma}^\prime[\CScPj]/d\hat{\sigma}^\prime[\COcSa]$ is compared to $-r^\prime(\chi_{c0})$. Unlike NRQCD factorization, there is a wide range of $r^\prime(\chi_{c0})$ in which $d\hat{\sigma}^\prime[\CScPj]/d\hat{\sigma}^\prime[\COcSa]$ is larger than $-r^\prime(\chi_{c0})$. In this value range, one can ensure that the cross section of $\chi_{cJ}$ production are positive within  the entire $p_T$ region. In the lower panel of Fig.~\ref{fig:sgfratio}, we show the differential cross sections when the central values in Eq.~\eqref{eq:SGFLDEM} are taken. Therefore, the negative cross section problem in NRQCD factorization can be resolved in the SGF framework.
%%%%%%%%%%%%%%%%%%%%%%%%%%%%%%%%%%%%%%%%%%%%%%%%%%%%%%%%%%%%%%%%%%%%%%%%%%%%%%%%%%%%%%%
%%%%%%%%%%%%%%%%%%%%%%%%%%%%%%%%%%%%%%%%%%%%%%%%%%%%%%%%%%%%%%%%%%%%%%%%%%%%%%%%%%%%%%%
%%%%%%%%%%%%%%%%%%%%%%%%%%%%%%%%%%%%%%%%%%%%%%%%%%%%%%%%%%%%%%%%%%%%%%%%%%%%%%%%%%%%%%%

\subsection{Production of $\psi(2S)$}\label{sec:psi2s}

For the $\psi(2S)$ production, the color-singlet contribution is neglectable as have been mentioned before. In NRQCD factorization, the determination of the LDMEs is complicated and involved. Similar to the previous
studies~\cite{Ma:2010yw,Ma:2010jj,Shao:2014yta}, we find that in the region where $12~\textrm{GeV} \leq p_T \leq 100~\textrm{GeV}$, the SDC of $\COcPz$ can be nicely decomposed into a linear combination of the SDCs of $\COcSa$ and $\COaSz$,
\begin{align}\label{eq:decom}
d\hat{\sigma}[\COcPz]= r_0 \frac{d\hat{\sigma}[\COaSz]}{m_c^2} + r_1 \frac{d\hat{\sigma}[\COcSa]}{m_c^2},
\end{align}
with $r_0=5.94$ and $r_1=-1.85$. Based on this relation, we are able to extract two linear combinations of the
three color-octet LDMEs within convincing precision,
\begin{subequations}\label{eq:combination}
\begin{align}
M_0^{\psi(2S)} &\equiv \langle \mathcal{O}^{\psi(2S)}(\COaSz)\rangle + r_0 \frac{\langle \mathcal{O}^{\psi(2S)}(\COcPz)\rangle}{m_c^2} , \\
M_1^{\psi(2S)} &\equiv \langle \mathcal{O}^{\psi(2S)}(\COcSa)\rangle + r_1 \frac{\langle \mathcal{O}^{\psi(2S)}(\COcPz)\rangle}{m_c^2} .
\end{align}
\end{subequations}
In SGF, we set $\bar{\Lambda}[\COcSa]=0.7~\textrm{GeV}$, $\bar{\Lambda}[\COaSz]=\bar{\Lambda}[\COcPz]=0.5~\textrm{GeV}\approx 0.7\bar{\Lambda}[\COcSa]$ for $\psi(2S)$ production. By convolving the SGDs with the short distance hard parts, we arrive at
\begin{align}
d\sigma(\psi(2S))=& d\hat{\sigma}^\prime[\COcSa]N^{\psi(2S)}[\COcSa]
  + d\hat{\sigma}^\prime[\COaSz]
  \nonumber\\ &\hspace{-1.5cm} \times N^{\psi(2S)}[\COaSz]
 + d\hat{\sigma}^\prime[\COcPz] \frac{N^{\psi(2S)}[\COcPz]}{m_c^2}.
\end{align}
Similar to the case of NRQCD factorization, in the region where $12~\textrm{GeV} \leq p_T \leq 100 ~\textrm{GeV}$, the coefficient $d\hat{\sigma}^\prime[\COcPz]$ can also be decomposed as
\begin{align}\label{eq:decom-sgf}
d\hat{\sigma}^\prime[\COcPz]= r_0^\prime \frac{d\hat{\sigma}^\prime[\COaSz]}{m_c^2} + r_1^\prime \frac{d\hat{\sigma}^\prime[\COcSa]}{m_c^2},
\end{align}
with $r_0^\prime=1.629$ and $r_1^\prime=-0.032$. And we construct following two parameters
\begin{subequations}\label{eq:combination-sgf}
\begin{align}
M_0^{\prime \psi(2S)} &\equiv N^{\psi(2S)}[\COaSz] + r_0^\prime \frac{N^{\psi(2S)}[\COcPz]}{m_c^2} , \\
M_1^{\prime \psi(2S)} &\equiv N^{\psi(2S)}[\COcSa] + r_1^\prime \frac{N^{\psi(2S)}[\COcPz]}{m_c^2} .
\end{align}
\end{subequations}

In this paper we are interested in the sign of the total contributions from $\COcSa$ and
$\COcPj$ channels. For this purpose, we need to determine the value range of quantities
\begin{align}
r(\psi(2S)) &\equiv\frac{\langle \mathcal{O}^{\psi(2S)}(\COcSa)\rangle}{\langle \mathcal{O}^{\psi(2S)}(\COcPz)\rangle/m_c^2},\nonumber\\
r^\prime(\psi(2S)) & \equiv \frac{N^{\psi(2S)}[\COcSa]}{N^{\psi(2S)}[\COcPz]/m_c^2}.
\end{align}
We assume that $N^{\psi(2S)}[\COcSa]$, $N^{\psi(2S)}[\COaSz]$, $N^{\psi(2S)}[\COcPz]$ and all of the three NRQCD color-octet LDMEs are positive.
Using the ATLAS and CMS yields data~\cite{ATLAS:2014zpz,CMS:2015lbl}, we extract
\begin{subequations}
\begin{align}
M_0^{\psi(2S)}=(2.92 \pm 0.49)\times 10^{-2} \textrm{GeV}^3,\\
M_1^{\psi(2S)}=(0.11 \pm 0.02)\times 10^{-2} \textrm{GeV}^3,
\end{align}
\end{subequations}
with $\chi^2/\textrm{d.o.f}=16.58/86$. And
\begin{subequations}
\begin{align}
M_0^{\prime\psi(2S)}=(7.37 \pm 1.51)\times 10^{-2} \textrm{GeV}^3,\\
M_1^{\prime\psi(2S)}=(0.85 \pm 0.11)\times 10^{-2} \textrm{GeV}^3,
\end{align}
\end{subequations}
with $\chi^2/\textrm{d.o.f}=15.21/86$. By varying $0\leq \langle \mathcal{O}^{\psi(2S)}(\COaSz)\rangle \leq M_0^{ \psi(2S)}$ and $0\leq N^{\psi(2S)}[\COaSz]  \leq M_0^{\prime \psi(2S)}$,
we further obtain the lower bounds of $r(\psi(2S))$ and $r^\prime(\psi(2S))$, i.e.,
\begin{align}
r(\psi(2S)) \geq 2.01, \quad\quad
r^\prime(\psi(2S)) \geq 0.17.
\end{align}

\begin{figure}[htb!]
 \includegraphics[width=0.8\linewidth]{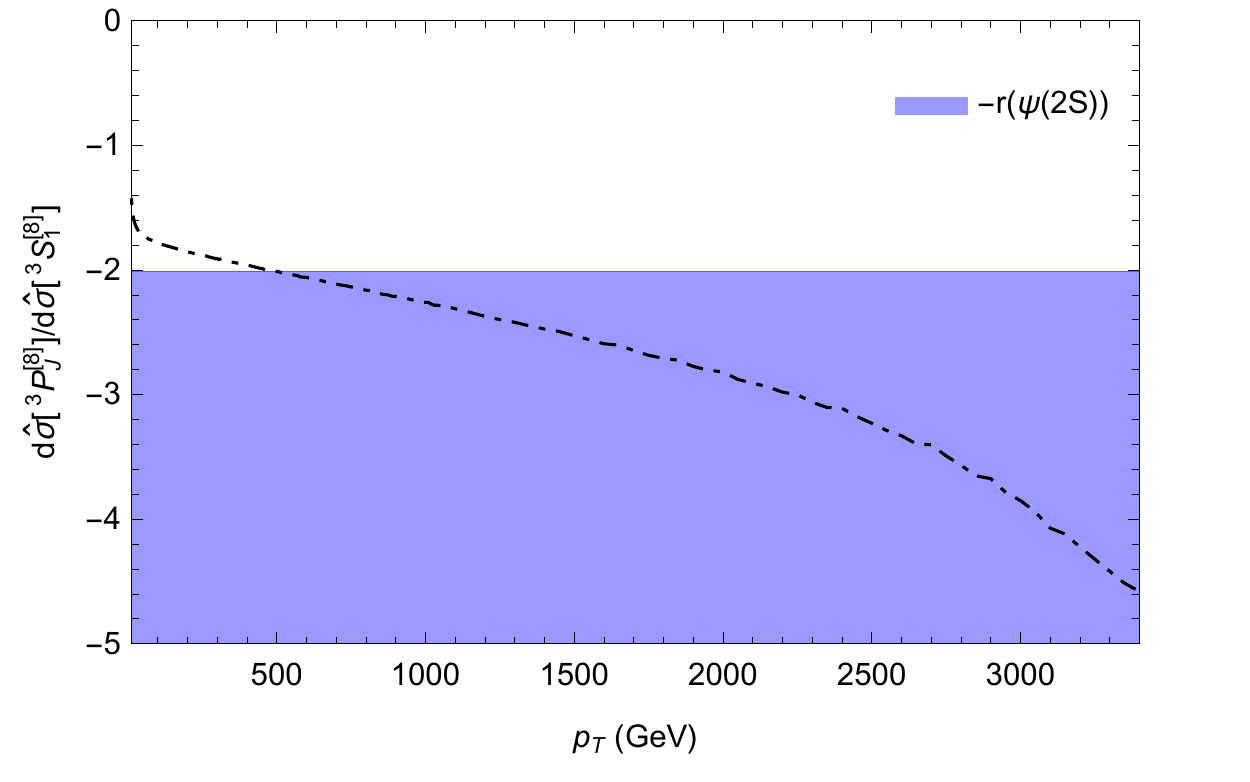}
 \includegraphics[width=0.8\linewidth]{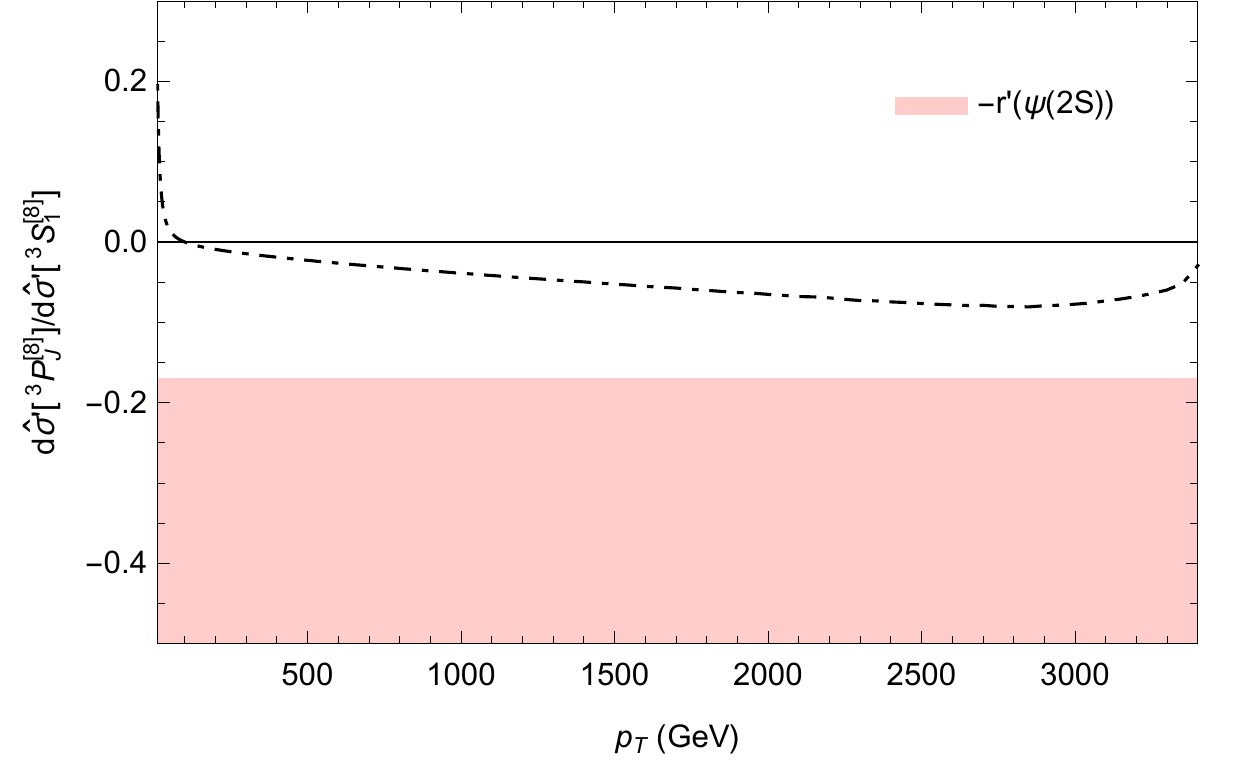}
 \caption{Upper panel: The comparison between the ratio $d\hat{\sigma}[\COcPz]/d\hat{\sigma}[\COcSa]$ and $-r(\psi(2S))$ in NRQCD factorization. Lower panel: The comparison between the ratio $d\hat{\sigma}^\prime[\COcPz]/d\hat{\sigma}^\prime[\COcSa]$ and $-r^\prime(\psi(2S))$ in SGF. The lower bounds of $-r(\psi(2S))$ and $-r^\prime(\psi(2S))$ go to $-\infty$. Here $\sqrt{s}=7~\rm{TeV}$, $|y|<0.75$.} \label{fig:psi2s}
\end{figure}

In Fig.~\ref{fig:psi2s}, we show the comparison between $r(\psi(2S))$ and $d\hat{\sigma}[\COcPz]/d\hat{\sigma}[\COcSa]$ in the upper panel , and show the comparison between $r^\prime(\psi(2S))$ and $d\hat{\sigma}^\prime[\COcPz]/d\hat{\sigma}^\prime[\COcSa]$ in the lower panel. We observe that both NRQCD factorization and SGF allow for a wide range of parameters that result in positive contributions of $\COcSa+\COcPj$ for the entire $p_T$ range. However, the SGF result is qualitatively better than that of NRQCD factorization due to that $d\hat{\sigma}^\prime[\COcPz]/d\hat{\sigma}^\prime[\COcSa]$ is larger than the upper bound of $-r^\prime(\psi(2S))$.
%%%%%%%%%%%%%%%%%%%%%%%%%%%%%%%%%%%%%%%%%%%%%%%%%%%%%%%%%%%%%%%%%%%%%%%%%%%%%%%%%%%%%%%

%\sect{Velocity scaling rules}

%Modification of velocity scaling rules in NRQCD.

%\sect{Perturbative calculation of LDMEs}

%This is possible because they are Sudakov-safe, although not IR-safe.

%%%%%%%%%%%%%%%%%%%%%%%%%%%%%%%%%%%%%%%%%%%%%%%%%
\section{Summary} \label{sec:summary}

In this paper, we study the hadroproduction of $\chi_{cJ}$ and $\psi(2S)$ using the SGF approach. Our general framework is the LP+NLP collinear factorization, with only the LO contribution considered. Our results show that the fit to experimental data in SGF is as good as that in NRQCD factorization. We confirm that the NRQCD predictions for $\chi_{cJ}$ production rates at the LHC turn negative at sufficiently large $p_T$. These negative cross sections originate from terms proportional to $1/(1-z)_+$ in the $\CScPj$ gluon FFs, which are remnants of the infrared subtraction in matching the $\CScPj$ SDCs and correspond to the leading-power order in emitted soft gluon momentum. At the threshold limit $z\to 1$, these terms contain nonperturbative effects. In SGF, these terms are factorized into the nonperturbative $\COcSa$ SGD, resulting in a distinct infrared subtraction scheme from NRQCD. We also calculate the short distance hard parts for relevant  FFs in SGF, and indeed, the $P$-wave short distance hard parts are free of the plus distributions.

In our phenomenological study, we employed a simple model for the nonperturbative SGD, which depends on three parameters: $b$, $\bar \Lambda$, and $N^H$. Currently, there is no first-principles method to determine these parameters. We fixed the parameter $b$ and varied $\bar{\Lambda}[\COcSa]$ and $\bar{\Lambda}[\CScPj]$. Based on our numerical results, we suggest the constraint $\bar{\Lambda}[\state{{3}}{P}{J}{1}] \geq 0.7\bar{\Lambda}[\COcSa]$, which indicates that the soft gluon emission in the hadronization of $Q\bar Q[\CScPj]$ to $\chi_{cJ}$ cannot be ignored in the current factorization framework. With an appropriate choice of $\bar{\Lambda}[\COcSa]$ and $\bar{\Lambda}[\CScPj]$, there are wide ranges of the parameters $N^{\chi_{c0}}[\COcSa]$ and $N^{\chi_{c0}}[\CScPz]$ that yield a positive cross section of $\chi_{cJ}$. Similarly, we also calculated the contributions of $\COcSa+\COcPj$ for $\psi(2S)$ production. We found that in both NRQCD factorization and SGF, there is a wide range of allowed parameters that yield positive contributions of $\COcSa+\COcPj$ at the entire $p_T$ region. Based on these results, we conclude that the negative cross section problem in NRQCD can be resolved in SGF.

%%%%%%%%%%%%%%%%%%%%%%%%%%%%%%%%%%%%%%%%%%%%%%%%%
\section*{Acknowledgments}
We thank Kuang-Ta Chao for many useful discussions. The work is supported
in part by the National Natural Science Foundation of China (Grants No. 12205124, No. 11875071, No. 11975029), the National Key Research and Development Program of
China under Contracts No. 2020YFA0406400.

{\bf Note added:} When our paper was being finalized, a preprint~\cite{Chung:2023ext} investigating a similar issue by applying the shape function formalism was appeared.

%%%%%%%%%%%%%%%%%%%%%%%%%%%%%%%%%%%%%%%%%%%%%%%%%

\appendix
%%%%%%%%%%%%%%%%%%%%%%%%%%%%%%%%%
%                                                   						
%                                   App I				
%  											   	
%%%%%%%%%%%%%%%%%%%%%%%%%%%%%%%%%
\section{The short distance hard parts in SGF}\label{app:hard-part-results}

In this Appendix, we provide the necessary details for obtaining the short distance hard parts in Eq.~\eqref{eq:FFHP}. In Eq.~\eqref{eq:SGD-1d}, the projection
operators $\mathcal {K}_{n}$ for $n=\state{3}{S}{1,S_z}{8}$, $\COaSz$ and $\state{3}{P}{J,J_z}{1,8}$ are given by~\cite{Ma:2017xno}
\begin{align}
 \mathcal {K}_{n}(rb^-) =& \frac{\sqrt{M_ {H}}}{M_ {H}+2 m_Q}\frac{ M_{H} + \slashed{p}}{2M_ {H}} \Gamma_n  \frac{M_{H} - \slashed{p}}{2M_{H}},
\end{align}
with
\begin{align}\label{eq:spinProj}
	\Gamma_n  =&
	\epsilon_{S_z}^\mu \gamma_\mu \mathcal {C}^{[8]}, &\text{for $n= \state{3}{S}{1,S_z}{8}$,}  \nonumber\\
	\Gamma_n  =&  \gamma_5 \mathcal {C}^{[8]}, &\text{for $n= \state{1}{S}{0}{8}$,}
 \\
	\Gamma_n  =&  \mathcal {E}_{J,J_z}^{\mu\nu} \gamma_\mu \Big( -\frac{i}{2}  \overleftrightarrow{D}_\nu \Big) \mathcal {C}^{[1,8]}, &\text{for $n= \state{3}{P}{J,J_z}{1,8}$.} \nonumber
\end{align}
The color operators $\mathcal {C}^{[c]}$ in above are defined as
\begin{subequations}
\begin{align}
\mathcal {C}^{[1]} =&\frac{{\bm 1}_c}{\sqrt{N_c}},\\
\mathcal {C}^{[8]} =& \sqrt{2}T^{ \bar a} \Phi_{l}(rb^-)_{\bar a a},
\end{align}
\end{subequations}
where ${\bm 1}_c$ and $T^{ \bar a}$ are the identity matrix and the generator of the fundamental (triplet) representation of $\textrm{SU}(3)$. The introduction of gauge link $\Phi_{l}(rb^-)_{\bar a   a}$ is to enable gauge invariance of SGDs. The gauge link is defined along the $l^\mu = (0, 1, \vec{0}_\perp)$ direction,
\begin{equation}\label{eq:gaugelink}
  \Phi_l(rb^{-})= \mathcal {P} \, \text{exp} \left[-i g_s
  \int_{0}^{\infty}\mathrm{d}\xi l\cdot A(rb^{-} + \xi l) \right] \, ,
\end{equation}
where $\mathcal {P}$ denotes path ordering, $A^{\mu}(x)$ is the matrix-valued gluon field in the adjoint representation: $[A^{\mu}(x)]_{ac} = i f^{abc} A^{\mu}_{b}(x)$.
In Eq.~\eqref{eq:spinProj}, $D_\mu$ is the gauge covariant derivative with $\overline\Psi \lrd_\mu \Psi =
\overline\Psi (D_\mu \Psi) -
(D_\mu \overline\Psi)\Psi$. And $\epsilon_{S_z}$, $\mathcal {E}_{J,J_z}$ are the polarization tensors for $\state{3}{S}{1,S_z}{8}$ state and $\state{{3}}{P}{J,J_z}{1,8}$ state.

According to Ref.~\cite{Ma:2015yka}, we use following projection
operators to sum over the polarizations of $\COcSa$ state and $\state{{3}}{P}{J,J_z}{1,8}$ state,
\begin{subequations}\label{eq:projection operator}
\begin{align}
&\mathbb{P}_{0}^{\beta\beta^\prime \sigma\sigma^\prime} \equiv
 \sum_{|J_z|=0}\mathcal {E}_{0,J_z}^{\beta\sigma} \mathcal {E}_{0,J_z}^{\ast \beta ^\prime\sigma^\prime}   = \frac{1}{d-1} \mathbb{P}^{\beta\sigma}\mathbb{P}^{\beta^\prime\sigma^\prime},\\
& \mathbb{P}_{1,T}^{\beta\beta^\prime \sigma\sigma^\prime} \equiv \sum_{|J_z|=1}\mathcal {E}_{1,J_z}^{\beta\sigma} \mathcal {E}_{1,J_z}^{\ast \beta ^\prime\sigma^\prime}   \nonumber\\&  =\frac{1}{2}
\Big(  \mathbb{P}_{\perp}^{\beta\beta^\prime}\mathbb{P}_{\parallel}^{\sigma\sigma^\prime} + \mathbb{P}_{\parallel}^{\beta\beta^\prime}\mathbb{P}_{\perp}^{\sigma\sigma^\prime}-
\mathbb{P}_{\perp}^{\beta\sigma^\prime}\mathbb{P}_{\parallel}^{\beta^\prime\sigma}-
\mathbb{P}_{\parallel}^{\beta\sigma^\prime}\mathbb{P}_{\perp}^{\beta^\prime\sigma}
\Big),\\
& \mathbb{P}_{1,L}^{\beta\beta^\prime \sigma\sigma^\prime} \equiv \sum_{|J_z|=0}\mathcal {E}_{1,J_z}^{\beta\sigma} \mathcal {E}_{1,J_z}^{\ast \beta ^\prime\sigma^\prime} \nonumber\\ &  = \frac{1}{2}
\Big(  \mathbb{P}_{\perp}^{\beta\beta^\prime}\mathbb{P}_{\perp}^{\sigma\sigma^\prime} -
\mathbb{P}_{\perp}^{\beta\sigma^\prime}\mathbb{P}_{\perp}^{\beta^\prime\sigma}
\Big),
\\
& \mathbb{P}_{2,TT}^{\beta\beta^\prime \sigma\sigma^\prime} \equiv \sum_{|J_z|=2}\mathcal {E}_{2,J_z}^{\beta\sigma} \mathcal {E}_{2,J_z}^{\ast \beta ^\prime\sigma^\prime} \nonumber\\ &  = \frac{1}{2}
\Big(  \mathbb{P}_{\perp}^{\beta\beta^\prime}\mathbb{P}_{\perp}^{\sigma\sigma^\prime} +
\mathbb{P}_{\perp}^{\beta\sigma^\prime}\mathbb{P}_{\perp}^{\beta^\prime\sigma}
\Big)
 - \frac{1}{d-2} \mathbb{P}_{\perp}^{\beta\sigma} \mathbb{P}_{\perp}^{\beta^\prime\sigma^\prime},\\
& \mathbb{P}_{2,T}^{\beta\beta^\prime \sigma\sigma^\prime} \equiv \sum_{|J_z|=1} \mathcal {E}_{2,J_z}^{\beta\sigma} \mathcal {E}_{2,J_z}^{\ast \beta ^\prime\sigma^\prime} \nonumber\\ &  = \frac{1}{2}
\Big(  \mathbb{P}_{\perp}^{\beta\beta^\prime}\mathbb{P}_{\parallel}^{\sigma\sigma^\prime} + \mathbb{P}_{\parallel}^{\beta\beta^\prime}\mathbb{P}_{\perp}^{\sigma\sigma^\prime}+
\mathbb{P}_{\perp}^{\beta\sigma^\prime}\mathbb{P}_{\parallel}^{\beta^\prime\sigma}+
\mathbb{P}_{\parallel}^{\beta\sigma^\prime}\mathbb{P}_{\perp}^{\beta^\prime\sigma}
\Big),\\
& \mathbb{P}_{2,L}^{\beta\beta^\prime \sigma\sigma^\prime} \equiv \sum_{|J_z|=0} \mathcal {E}_{2,J_z}^{\beta\sigma} \mathcal {E}_{2,J_z}^{\ast \beta ^\prime\sigma^\prime} \nonumber\\ &  = \frac{d-2}{d-1}
\Big(
\mathbb{P}_{\parallel}^{\beta\sigma} - \frac{1}{d-2} \mathbb{P}_{\perp}^{\beta\sigma}
\Big)
\Big(
\mathbb{P}_{\parallel}^{\beta^\prime\sigma^\prime} - \frac{1}{d-2} \mathbb{P}_{\perp}^{\beta^\prime\sigma^\prime}
\Big).
 \end{align}
 \end{subequations}
 Where
\begin{subequations}
\begin{align}
&\mathbb{P}_{\perp}^{\alpha\alpha^\prime}\equiv \sum_{|S_z|=1}\epsilon_{S_z}^\alpha\epsilon_{S_z}^{\ast \alpha^\prime}   \nonumber\\
& = -g^{\alpha\alpha^\prime} + \frac{p^\alpha l^{\alpha^\prime} + p^{\alpha^\prime} l^\alpha}{ p\cdot l} - \frac{p^2}{(p\cdot l)^2}l^{\alpha}l^{\alpha^\prime},\\
&\mathbb{P}_{\parallel}^{\alpha\alpha^\prime}\equiv \sum_{|S_z|=0}\epsilon_{S_z}^\alpha\epsilon_{S_z}^{\ast \alpha^\prime} \nonumber\\
&= \frac{p^{\alpha}p^{\alpha^\prime}}{p^2} - \frac{p^\alpha l^{\alpha^\prime} + p^{\alpha^\prime} l^\alpha}{ p\cdot l} + \frac{p^2}{(p\cdot l)^2}l^{\alpha}l^{\alpha^\prime}, \\
& \mathbb{P}^{\alpha\alpha^\prime} \equiv \sum_{S_z}\epsilon_{S_z}^\alpha\epsilon_{S_z}^{\ast \alpha^\prime}
= -g^{\alpha\alpha^\prime} + \frac{p^{\alpha}p^{\alpha^\prime}}{p^2}.
\end{align}
\end{subequations}

Following the matching procedure, to determine the short distance hard part in Eq.~\eqref{eq:SGF-form}, we replace the quarkonium $H$ by a on-shell $Q\bar{Q}$ pair with certain quantum number $n$ and momenta
\begin{align}
p_Q= \frac{1}{2}p +q ,  \quad \quad p_{\bar Q}= \frac{1}{2}p -q.
\end{align}
where $q$ is half of the relative momentum of the $Q\bar Q$ pair. On-shell conditions $p_Q^2= p_{\bar Q}^2=m_Q^2$ result in
\begin{align}
p \cdot q= 0 , \quad \quad q^2=m_Q^2-p^2/4.
\end{align}
To project the final-state $Q \bar Q$ pair to the state $n$, we replace spinors of the $Q \bar Q$ by the projector~\cite{Ma:2017xno}
\begin{align}\label{eq:projector}
& \int \frac{\mathrm{d}^{d-2}\Omega}{N_\Omega} \frac{2}{\sqrt{M_H}(M_H+2m_Q)} ( \slashed{p}_{\bar Q} - m_Q ) \nonumber\\
& \times  \frac{M_H - \slashed{p}}{2M_H}  \tilde{\Gamma}_n
\frac{M_H + \slashed{p}}{2M_H} (\slashed{p}_{Q} +m_Q ),
\end{align}
here $\Omega$ is the solid angle of relative momentum $\textbf{q}$ in the $Q \bar Q$ rest frame, and $N_\Omega$ is given by
\begin{align}
N_\Omega = \int \mathrm{d}^{d-2}\Omega.
\end{align}
For different states $n$, the operators $\tilde{\Gamma}_n$ are given by
\begin{align}
	\tilde{\Gamma}_n  =&
	\epsilon_{S_z}^{\ast \mu} \gamma_\mu \tilde{\mathcal {C}}^{[8]}, &\text{for $n= \state{3}{S}{1,S_z}{8}$,}  \nonumber\\
	\tilde{\Gamma}_n  =&  \gamma_5 \tilde{\mathcal {C}}^{[8]}, &\text{for $n= \state{1}{S}{0}{8}$,}
 \\
	\tilde{\Gamma}_n  =&  \frac{(d-1) q_\alpha }{ \textbf{q}^2}  \mathcal {E}_{J,J_z}^{\ast \alpha \mu}\gamma_\mu \tilde{\mathcal {C}}^{[1,8]}, &\text{for $n= \state{3}{P}{J,J_z}{1,8}$,} \nonumber
\end{align}
where $\textbf{q}^2=-q^2$, and
\begin{subequations}
\begin{align}
\tilde{\mathcal {C}}^{[1]} =&\frac{{\bm 1}_c}{\sqrt{N_c}},\\
\tilde{\mathcal {C}}^{[8]} =& \sqrt{\frac{2}{N_c^2-1}}T^{  a} .
\end{align}
\end{subequations}

 We first consider the single parton FFs. Basing on Eq.~\eqref{eq:SGF-form}, we can derive
following matching relations for the short distance hard parts at LO in $\alpha_s$
\begin{subequations}\label{eq:matching-relation-2}
\begin{align}
&\hat{D}_{g \to Q\bar{Q}[\state{{3}}{S}{1,\lambda}{8}]}^{LO,(0)}(z; M_H,\mu_0,\mu_\Lambda)
 \nonumber\\
 =& D^{LO}_{g \to Q\bar{Q}[\state{{3}}{S}{1,\lambda}{8}]}(z; M_H,m_Q,\mu_0)\Big \vert_{m_Q^2 = M_H^2/4}, \\
 &\hat{D}_{g \to Q\bar{Q}[\COaSz]}^{LO,(0)}(z; M_H,\mu_0,\mu_\Lambda)
 \nonumber\\
 =& D^{LO}_{g \to Q\bar{Q}[\COaSz]}(z; M_H,m_Q,\mu_0)\Big \vert_{m_Q^2 = M_H^2/4}, \\
&\hat{D}_{g \to Q\bar{Q}[\state{{3}}{P}{J,\lambda}{1,8}]}^{LO,(0)}(z; M_H,\mu_0,\mu_\Lambda)
 \nonumber\\
=&  \Big [D^{LO}_{g \to Q\bar{Q}[\state{{3}}{P}{J,\lambda}{1,8}]}(z; M_H,m_Q,\mu_0)
\nonumber\\
& - \sum_{\lambda^\prime=T,L} \int \frac{\mathrm{d}x}{x^2}
 \hat{D}^{LO}_{ g \to Q\bar{Q}[\state{{3}}{S}{1,\lambda^\prime}{8}]}(\hat{z}; M_H/x, m_Q,\mu_0, \mu_\Lambda)
\nonumber\\
&\times
 F^{LO}_{[\state{{3}}{S}{1,\lambda^\prime}{8}] \to Q\bar{Q}[\state{{3}}{P}{J,\lambda}{1,8}]}(x;M_H,m_Q,\mu_\Lambda) \Big] \Big \vert_{m_Q^2 = M_H^2/4} .
\end{align}
\end{subequations}
Here we have used~\cite{Ma:2017xno}
\begin{align}
F^{LO}_{[n] \to Q\bar{Q}[n]}(x;M_H,m_Q,\mu_\Lambda)=\delta(1-z),
\end{align}
for $n=\state{3}{S}{1,S_z}{8}$, $\COaSz$ and $\state{3}{P}{J,J_z}{1,8}$.
The short distance hard part for $\state{{3}}{S}{1,\lambda}{8}$ can be taken from~\cite{Chen:2021hzo}, which read as
 \begin{subequations}\label{eq:FFHP-3s18-1}
\begin{align}
        \hat{D}^{LO}_{ g \to Q\bar Q[\state{{3}}{S}{1,T}{8}] }(z; M_H, m_Q,\mu_0, \mu_\Lambda)=& \frac{\pi \alpha_s \mu_c^{2\epsilon}}{(N_c^2-1)}\frac{8}{M_H^3}
        \nonumber\\
        & \hspace{-4.5cm} \times \Big ( \frac{2- 2\epsilon }{3-2\epsilon}+ \frac{ 2  m_Q}{(3-2\epsilon) M_H}\Big)^2 \delta(1-  z), \\
        \hat{D}^{LO}_{ g \to Q\bar Q[\state{{3}}{S}{1,L}{8}] }(z; M_H, m_Q,\mu_0, \mu_\Lambda)=& 0,
\end{align}
\end{subequations}
here $\mu_c$ is the dimensional regularization scale. Expanding $m_Q^2$ around $M_H^2/4$,
we then obtain
\begin{subequations}\label{eq:FFHP-3s18}
\begin{align}
 \hat{D}_{g \to Q\bar Q[\state{{3}}{S}{1,T}{8}]}^{LO,(0)}( z,M_H, \mu_0, \mu_\Lambda) =&  \frac{\pi \alpha_s}{(N_c^2-1)}\frac{8}{M_H^3}  \delta(1-z),
\\
\hat{D}_{g \to Q\bar Q[\state{{3}}{S}{1,L}{8}]}^{LO,(0)}( z,M_H, \mu_0, \mu_\Lambda) =&  0.
\end{align}
\end{subequations}
The calculation of $\hat{D}_{g \to Q\bar{Q}[\COaSz]}^{LO,(0)}$ is straightforward. Using the results of perturbative FFs listed in~\cite{Ma:2013yla}, we derive
\begin{align}
\hat{D}_{g \to Q\bar Q[\state{{1}}{S}{0}{8}]}^{LO,(0)}( z,M_H, \mu_0, \mu_\Lambda) =& \frac{8\alpha_s^2}{M_H^3} \frac{N_c^2-4}{2N_c(N_c^2-1)}
\nonumber\\
& \hspace{-4cm} \times
 \Big[(1-z)\ln[1-z] -z^2 + \frac{3}{2}z \Big].
\end{align}

The computation of perturbative FFs $D^{LO}_{g \to Q\bar{Q}[\state{{3}}{P}{J,\lambda}{1,8}]}$ in Eq.~\eqref{eq:matching-relation-2} has been well studied in the literature (e.g., Ref.~\cite{Zhang:2020atv} ). Using the projection operators in Eq.~\eqref{eq:projection operator}, we obtain
\begin{subequations}\label{eq:perturbative-FF}
\begin{align}
& D^{LO}_{g \to Q\bar{Q}[\state{{3}}{P}{0}{1}]}(z;M_H,m_Q,\mu_0) \nonumber\\
=&\frac{32\alpha_s^2\mu_c^{2\epsilon}}{M_H^5 N_c} \frac{2}{9} \Big[ \Big( -\frac{ 1}{ \epsilon_{\textrm{IR}}  }- \frac{1}{6} - \ln \frac{4\pi\mu_c^2 e^{-\gamma_E}}{M_H^2} \Big) \delta(1-z)
         \nonumber\\
        & + \frac{z \left(26 z^2-111 z+93\right)}{4}\frac{1}{(1-z)_+} + \frac{9 \left(5 - 3 z\right)  }{2 }
        \nonumber\\ & \times \ln (1-z) \Big]
         + \mathcal {O}(\boldsymbol{q}^2), \\
& D^{LO}_{g \to Q\bar{Q}[\state{{3}}{P}{1,T}{1}]}(z;M_H,m_Q,\mu_0) \nonumber\\
=& \frac{32
        \alpha_s^2\mu_c^{2\epsilon}}{M_H^5 N_c} \frac{1}{3} \Big[ \Big(-\frac{ 1}{ \epsilon_{\textrm{IR}}  } - \ln \frac{4\pi\mu_c^2 e^{-\gamma_E}}{M_H^2} \Big) \delta(1-z)
         \nonumber\\
        &
        + 2 z\left( z^2- z+1\right)\frac{1}{  (1-z)_+} \Big] + \mathcal {O}(\boldsymbol{q}^2), \\
& D^{LO}_{g \to Q\bar{Q}[\state{{3}}{P}{1,L}{1}]}(z;M_H,m_Q,\mu_0) \nonumber\\
=&
        \frac{32\alpha_s^2\mu_c^{2\epsilon}}{M_H^5 N_c} \frac{1}{3} \Big[ \Big( -\frac{ 1}{ \epsilon_{\textrm{IR}}  } + \frac{5}{2} - \ln \frac{4\pi\mu_c^2 e^{-\gamma_E}}{M_H^2} \Big)\delta(1-z)
         \nonumber\\
        &
        + z \left(2 z^2- z+1\right) \frac{ 1}{ (1-z)_+}   \Big] + \mathcal {O}(\boldsymbol{q}^2), \\
&D^{LO}_{g \to Q\bar{Q}[\state{{3}}{P}{2,TT}{1}]}(z;M_H,m_Q,\mu_0) \nonumber\\
=&
        \frac{32\alpha_s^2\mu_c^{2\epsilon}}{M_H^5 N_c} \frac{2}{3z^4} \Big[ \Big( -\frac{ 1}{ \epsilon_{\textrm{IR}}  } +2 - \ln \frac{4\pi\mu_c^2 e^{-\gamma_E}}{M_H^2} \Big)\delta(1-z)
         \nonumber\\
        &
         + 2z^4 \frac{ 1 }{ (1-z)_+}
        -2 z (z^5-7 z^4+26 z^3-37 z^2+24 z \nonumber\\
        & -12)  - 6 \left(z^5-6 z^4+14 z^3-16 z^2+10 z-4\right)
         \nonumber\\ & \times \ln(1-z)  \Big] + \mathcal {O}(\boldsymbol{q}^2), \\
& D^{LO}_{g \to Q\bar{Q}[\state{{3}}{P}{2,T}{1}]}(z;M_H,m_Q,\mu_0) \nonumber\\
=&
        \frac{32\alpha_s^2\mu_c^{2\epsilon}}{M_H^5 N_c} \frac{1}{3z^4} \Big[ \Big( -\frac{ 1}{ \epsilon_{\textrm{IR}}  } - \ln \frac{4\pi\mu_c^2 e^{-\gamma_E}}{M_H^2} \Big) \delta(1-z)
        \nonumber\\
        & + 2z^4 \frac{ 1 }{ (1-z)_+}
         - 2z (z^5+4 z^4-55 z^3+152 z^2
         \nonumber\\
        & -192 z+96)  - 48 (z^4-5 z^3+10 z^2-10z+4)
        \nonumber\\ & \times \ln(1-z)  \Big] + \mathcal {O}(\boldsymbol{q}^2), \\
& D^{LO}_{g \to Q\bar{Q}(\state{{3}}{P}{2,L}{1})}(z;M_H,m_Q,\mu_0) \nonumber\\
 =&
        \frac{32\alpha_s^2\mu_c^{2\epsilon}}{M_H^5 N_c} \frac{1}{9z^4} \Big[ \Big( -\frac{ 1}{ \epsilon_{\textrm{IR}}  }- \frac{7}{6} - \ln \frac{4\pi\mu_c^2 e^{-\gamma_E}}{M_H^2} \Big)\delta(1-z)
         \nonumber\\
        &
        + 2z^4 \frac{ 1 }{ (1-z)_+}  -z (2 z^5+5 z^4+38 z^3-468 z^2
        \nonumber\\
        & +864 z-432) - 216(z - 2)(z - 1)^2
         \ln(1-z)  \Big]
         \nonumber\\ & + \mathcal {O}(\boldsymbol{q}^2),\\
& D^{LO}_{g \to Q\bar{Q}(\state{{3}}{P}{J,\lambda}{8})}(z;M_H,m_Q,\mu_0) \nonumber\\ =& \frac{N_c^2-4}{2(N_c^2-1)} D^{LO}_{g \to Q\bar{Q}(\state{{3}}{P}{J,\lambda}{1})}(z;M_H,m_Q,\mu_0).
\end{align}
\end{subequations}

\begin{figure}[htb!]
 \includegraphics[width=0.8\linewidth]{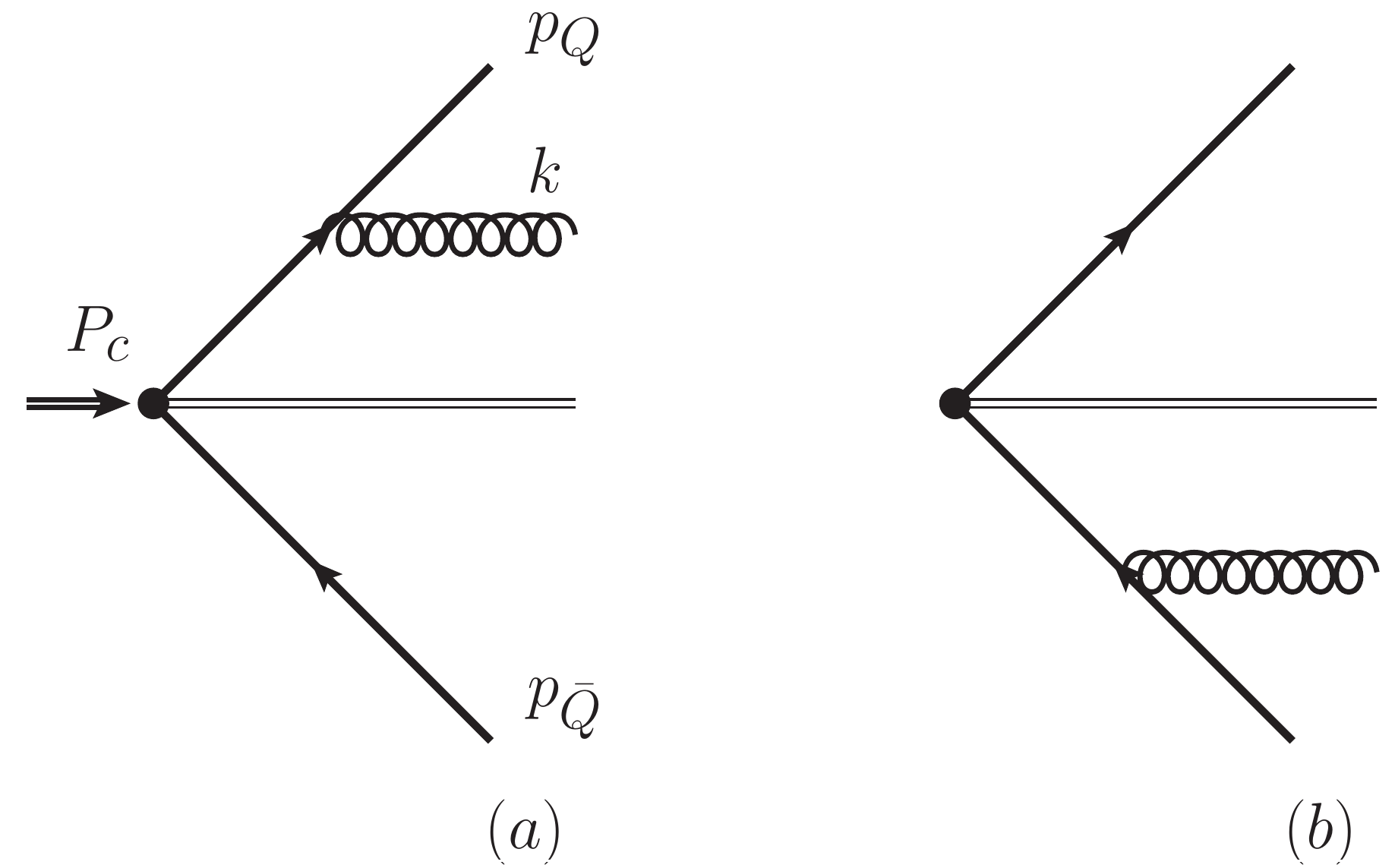}
 \caption{Feynman diagrams for the SGDs $F_{[\state{{3}}{S}{1,T}{1}] \to Q\bar Q[\state{{3}}{P}{J,\lambda}{1,8}]}$ at LO. The double solid line represents the gauge link along $l$ direction.} \label{fig:feynman-diagram}
 \vspace*{0.cm}
\end{figure}

We now turn to the calculation of SGDs in Eq.~\eqref{eq:matching-relation-2}. At LO, the diagrams for $F_{[\state{{3}}{S}{1,\lambda^\prime}{1}] \to Q\bar{Q}[\state{{3}}{P}{J,\lambda}{1,8}]}$ are shown in Fig.~\ref{fig:feynman-diagram}, here we only need to consider the case of $\lambda^\prime=T$.
%%%%%%%%%%%%%%%%%%%%%%%%%%%%%%%%%%%%%%%%%%%%%%%%%%%%%%%%%%%%%%%%%%%%%%%%%%%%%%%%%%%%%%%
%%%%%%%%%%%%%%%%%%%%%%%%%%%%%%%%%%%%%%%%%%%%%%%%%%%%%%%%%%%%%%%%%%%%%%%%%%%%%%%%%%%%%%%
According to the definition Eq.~\eqref{eq:SGD-1d}, we expand the amplitudes in terms of the soft momentum $k$, and keeping only the leading terms in the expansion. In addition, according to Eq.~\eqref{eq:matching-relation-2}, we can expand $m_Q^2$
in the amplitudes around $M_H^2/4$ and
neglect the terms of $\mathcal {O}(\boldsymbol{q}^2)$ before performing phase space integration. Thus we derive
\begin{align}\label{eq:SGD-SP}
& F_{[\state{{3}}{S}{1,T}{1}] \to Q\bar Q[\state{{3}}{P}{J,\lambda}{1}]}^{LO}(x,M_H,m_Q,\mu_\Lambda) \nonumber\\
 =& \frac{p^+}{d-2} \int \frac{\mathrm{d}^d P_c}{(2\pi)^d}\frac{\mathrm{d}^dk}{(2\pi)^d} \delta(P_c^+-\frac{p^+}{x})
(2\pi)^d
 \delta^{d}(P_c-p-k)
\nonumber\\
&
\times  2\pi \delta(k^2)\theta(k^+)
 \mathcal {M}_{\alpha\beta\sigma\rho} \mathcal {M}^{\ast}_{ \alpha^\prime \beta^\prime \sigma^\prime \rho^\prime} \mathbb{P}_{\perp}^{\alpha\alpha^\prime} \mathbb{P}_{J,\lambda}^{\beta\beta^\prime \sigma\sigma^\prime}(-g^{\rho\rho^\prime}) ,
\end{align}
where the amplitude $\mathcal {M}$ is given by
\begin{align}
\mathcal {M}_{\alpha\beta\sigma\rho} =& \frac{\mathrm{d}}{\mathrm{d}q^{\sigma}}[\mathcal {A}_{\alpha\beta\rho}^{(1)} +\mathcal {A}_{\alpha\beta\rho}^{(2)}] \vert_{q=0} + \mathcal {O}(\boldsymbol{q}^2), \nonumber\\
\mathcal {A}_{\alpha\beta\rho}^{(1)} =& g_s\textrm{Tr} \Big[ T^a \Pi_{\alpha}^{  b}
 \tilde{\Pi}_{ \beta} \Big]
\frac{(p/2+q)_\rho}{(p/2+q)\cdot k + i\varepsilon},
\nonumber\\
\mathcal {A}_{\alpha\beta\rho}^{(2)} =& g_s\textrm{Tr} \Big[  \Pi_{\alpha}^{ b} T^a
 \tilde{\Pi}_{ \beta} \Big]\frac{-(p/2-q)_\rho}{(p/2-q) \cdot k + i\varepsilon},
\end{align}
with
 \begin{align}
 \Pi_{\alpha}^{ b} =&  \frac{\sqrt{M_H}}{M_H+2m_Q}   \frac{M_H + \slashed{p}}{2M_H} (\sqrt{2}T^b \gamma_\alpha )
\frac{M_H - \slashed{p}}{2M_H},
\nonumber\\
 \tilde{\Pi}_\beta=& \frac{2}{\sqrt{M_H}(M_H+2m_Q)} ( \slashed{p}_{\bar Q} - m_Q )  \frac{M_H - \slashed{p}}{2M_H}
 \nonumber\\
 &\times \Big(\frac{{\bm 1}_c}{\sqrt{N_c}} \gamma_\beta \Big)
\frac{M_H + \slashed{p}}{2M_H} (\slashed{p}_{Q} +m_Q ).
\end{align}
Performing the $k$-integration, we obtain
\begin{subequations}\label{eq:SGD-result}
\begin{align}
& F_{[\state{{3}}{S}{1,T}{1}] \to Q\bar Q[\state{{3}}{P}{0}{1}]}^{LO}(x,M_H,m_Q,\mu_\Lambda) \nonumber\\
=&  \frac{\alpha_s }{ M_H^2 \pi }\frac{N_c^2-1}{N_c}\frac{8}{9}
        \Big[ \Big( -\frac{1}{\epsilon_{ \textrm{IR} }}-  \ln \frac{ 4\pi \mu_c^2 e^{ - \gamma_E}}{M_H^2} - \frac{1}{6} \Big)
        \nonumber\\
        & \times \delta(1-x)
        + 2x \frac{1}{(1-x)_+} \Big]
        + \mathcal {O}(\boldsymbol{q}^2), \\
& F_{[\state{{3}}{S}{1,T}{1}] \to Q\bar
        Q[\state{{3}}{P}{1,T}{1}]}^{LO}(x,M_H,m_Q,\mu_\Lambda) \nonumber\\
=&  \frac{\alpha_s }{ M_H^2 \pi }\frac{N_c^2-1}{N_c}\frac{4}{3}
        \Big[ \Big( -\frac{1}{\epsilon_{ \textrm{IR} }} -  \ln \frac{ 4\pi \mu_c^2 e^{ - \gamma_E}}{M_H^2}   \Big)
        \nonumber\\
        & \times
        \delta(1-x)
        + 2x \frac{1}{(1-x)_+} \Big]
        + \mathcal {O}(\boldsymbol{q}^2),\\
& F_{[\state{{3}}{S}{1,T}{1}] \to Q\bar
        Q[\state{{3}}{P}{1,L}{1}]}^{LO}(x,M_H,m_Q,\mu_\Lambda) \nonumber\\
=&  \frac{\alpha_s }{ M_H^2 \pi }\frac{N_c^2-1}{N_c}\frac{4}{3}
        \Big[ \Big( -\frac{1}{\epsilon_{ \textrm{IR} }}-  \ln \frac{ 4\pi \mu_c^2 e^{ - \gamma_E}}{M_H^2} + \frac{5}{2}  \Big)
        \nonumber\\
        & \times
        \delta(1-x)
        + 2x \frac{1}{(1-x)_+} \Big]
         + \mathcal {O}(\boldsymbol{q}^2) ,\\
& F_{[\state{{3}}{S}{1,T}{1}] \to Q\bar
        Q[\state{{3}}{P}{2,TT}{1}]}^{LO}(x,M_H,m_Q,\mu_\Lambda) \nonumber\\
=&  \frac{\alpha_s }{ M_H^2 \pi }\frac{N_c^2-1}{N_c}\frac{8}{3}
        \Big[ \Big( -\frac{1}{\epsilon_{ \textrm{IR} }}-  \ln \frac{ 4\pi \mu_c^2 e^{ - \gamma_E}}{M_H^2} + 2  \Big)
        \nonumber\\
        & \times
        \delta(1-x)
        + 2x \frac{1}{(1-x)_+} \Big]
        + \mathcal {O}(\boldsymbol{q}^2)  ,\\
& F_{[\state{{3}}{S}{1,T}{1}] \to Q\bar
        Q[\state{{3}}{P}{2,T}{1}]}^{LO}(x,M_H,m_Q,\mu_\Lambda) \nonumber\\
=&  \frac{\alpha_s }{ M_H^2 \pi }\frac{N_c^2-1}{N_c}\frac{4}{3}
        \Big[ \Big( -\frac{1}{\epsilon_{ \textrm{IR} }}-  \ln \frac{ 4\pi \mu_c^2 e^{ - \gamma_E}}{M_H^2}    \Big)
        \nonumber\\
        &\times
        \delta(1-x)
        + 2x \frac{1}{(1-x)_+} \Big]
         + \mathcal {O}(\boldsymbol{q}^2)  ,\\
& F_{[\state{{3}}{S}{1,T}{1}] \to Q\bar
        Q[\state{{3}}{P}{2,L}{1}]}^{LO}(x,M_H,m_Q,\mu_\Lambda) \nonumber\\
=&  \frac{\alpha_s }{ M_H^2 \pi }\frac{N_c^2-1}{N_c}\frac{4}{9}
        \Big[ \Big( -\frac{1}{\epsilon_{ \textrm{IR} }}-  \ln \frac{ 4\pi \mu_c^2 e^{ - \gamma_E}}{M_H^2} - \frac{7}{6}  \Big)
        \nonumber\\
        &\times \delta(1-x) +  2x\frac{1}{(1-x)_+} \Big]
         + \mathcal {O}(\boldsymbol{q}^2).
\end{align}
\end{subequations}
The SGDs $F_{[\state{{3}}{S}{1,T}{1}] \to Q\bar Q[\state{{3}}{P}{J,\lambda}{8}]}$ can be calculated similarly, with results given as
 \begin{align}\label{eq:color-octet-SGD}
& F_{[\state{{3}}{S}{1,T}{1}] \to Q\bar
        Q[\state{{3}}{P}{J,\lambda}{8}]}^{LO}(x,M_H,m_Q,\mu_\Lambda) \nonumber\\
=& \frac{N_c^2-4}{2(N_c^2-1)} F_{[\state{{3}}{S}{1,T}{1}] \to Q\bar
        Q[\state{{3}}{P}{J,\lambda}{1}]}^{LO}(x,M_H,m_Q,\mu_\Lambda).
\end{align}

Substituting Eqs.~\eqref{eq:color-octet-SGD},~\eqref{eq:SGD-result},~\eqref{eq:perturbative-FF} and \eqref{eq:FFHP-3s18-1} into Eq.~\eqref{eq:matching-relation-2}, we can obtain the $P$-wave short distance hard parts that given in Eq.~\eqref{eq:FFHP}. We find that both the infrared divergences and the terms proportional to $1/(1-z)_+$ in the perturbative FFs are correctly subtracted by the SGDs.

Finally, for the double parton FFs, we have following matching relation
 \begin{align}\label{eq:matching-relation-3}
&\hat{\cal{D}}^{LO,(0)}_{ [Q\bar{Q}(\kappa)] \to [Q\bar{Q}[n]] }(z, \zeta,\zeta^\prime; M_H,\mu_0,\mu_\Lambda)
\nonumber\\
=&   {\cal D}^{LO}_{[Q\bar{Q}(\kappa)]\to [Q\bar{Q}[n]]}(z,\zeta,\zeta^\prime; M_H,m_Q,\mu_0) \vert_{m_Q^2=M_H^2/4}.
\end{align}
Using the results of perturbative double parton FFs calculated in~\cite{Ma:2013yla,Ma:2014eja,Ma:2015yka}, we immediately obtain the results in Eq.~\eqref{eq:FFHP}.

%%%%%%%%%%%%%%%%%%%%%%%%%%%%%%%%%%%%%%%%%%%%%%%%%

% references
\providecommand{\href}[2]{#2}\begingroup\raggedright\endgroup

%%%%%%%%%%%%%%%%%%%%%%%%%%%%%%%%%%%%%%%%%%%%%%%%%
\end{document}